\documentclass{egpubl}
\usepackage[T1]{fontenc}
\usepackage{dfadobe}  

\usepackage{cite}  
\BibtexOrBiblatex
\electronicVersion
\PrintedOrElectronic
\ifpdf \usepackage[pdftex]{graphicx} \pdfcompresslevel=9
\else \usepackage[dvips]{graphicx} \fi

\usepackage{egweblnk} 

\usepackage[ruled,vlined]{algorithm2e}
\usepackage{amsmath}
\usepackage{cleveref}
\usepackage{listing}
\usepackage{xcolor}
\usepackage{booktabs}
\usepackage{multirow}
\usepackage{makecell}

\title[Minimizing Ray Tracing Memory Traffic]%
      {Minimizing Ray Tracing Memory Traffic through Quantized Structures and Ray Stream Tracing}

\author[M. Grauer \& J. Hanika \& C. Dachsbacher]{Moritz Grauer\orcid{0009-0004-6905-7301}, Johannes Hanika\orcid{0000-0002-7648-1782}, Carsten Dachsbacher\orcid{0000-0003-4690-3574}\\Karlsruhe Institute of Technology, Germany}

\begin{document}


\maketitle
\begin{abstract}
Memory bandwidth constraints continue to be a significant limiting factor in ray
tracing performance, particularly as scene complexity grows and computational
capabilities outpace memory access speeds. This paper presents a
memory-efficient ray tracing methodology that integrates compressed data
structures with ray stream techniques to reduce memory traffic. The approach
implements compressed BVH and triangle representations to minimize acceleration
structure size in combination with ray stream tracing to reduce traversal stack
memory traffic. The technique employs fixed-point arithmetic for intersection
tests for prospective hardware with tailored integer operations. Despite using
reduced precision, geometric holes are avoided by leveraging fixed-point
arithmetic instead of encountering the floating-point rounding errors common in
traditional approaches.
Quantitative analysis demonstrates significant memory traffic
reduction across various scene complexities and BVH configurations. The
presented 8-wide BVH ray stream implementation reduces memory traffic to only
18\% of traditional approaches by using 8-bit quantization for box and triangle
coordinates and directly ray tracing these quantized structures. These
reductions are especially beneficial for bandwidth-constrained hardware
environments such as mobile devices. This integrated approach addresses both
memory bandwidth limitations and numerical precision challenges inherent to
modern ray tracing applications.

\end{abstract}  
\section{Introduction}

Ray tracing is a fundamental technique in computer graphics, enabling the creation of photorealistic images by simulating the physical behavior of light. While computational power has increased dramatically over the past decades, memory bandwidth constraints remain a significant bottleneck for ray tracing performance. This is particularly evident as scene complexity grows with modern applications frequently requiring millions of primitives. The disparity between computational capabilities and memory access speeds has become increasingly pronounced, with compute capability advancing faster than memory bandwidth.

In ray tracing, the primary operations---traversing acceleration structures and intersecting geometric primitives---generate substantial memory traffic. Acceleration structures like Bounding Volume Hierarchies (BVHs) are essential for efficient ray traversal, but their memory footprint and the resulting bandwidth requirements can severely limit performance, especially for incoherent rays typical in global illumination algorithms.

We present a memory-efficient ray tracing methodology that addresses these bandwidth limitations through two complementary approaches: data compression and traversal optimization. Our key insight is that by jointly compressing the acceleration structure and geometry data, while simultaneously reducing traversal stack memory traffic, we can achieve significant bandwidth reductions with minimal precision loss.

Our approach makes the following contributions:
\begin{itemize}
    \item A unified quantization scheme for both BVH nodes and triangle geometry that uses 8-bit fixed-point representations within local coordinate systems.
    \item Direct ray tracing on compressed structures using fixed-point
    arithmetic, eliminating decompression overhead and enabling custom tailored
    hardware units.
    \item Integration of ray stream tracing techniques with wide BVHs (2, 4, and
    8-wide) to reduce traversal stack traffic and enable SIMD processing.
    \item A careful analysis of precision and performance trade-offs, demonstrating that our fixed-point representation avoids geometric holes common in floating-point approaches.
\end{itemize}

Our results show that this integrated approach reduces memory traffic to only 18\% of traditional methods for an 8-wide BVH with 8-bit quantization, while maintaining visual fidelity. 
By addressing the memory bandwidth bottleneck, our work aims to make ray tracing more practical for a wider range of applications and hardware platforms.

\section{Related Work}

Ray tracing research has evolved significantly over the decades, with numerous approaches targeting performance, memory efficiency, and quality. Our work builds upon several key research areas: ray coherence exploitation, memory-efficient acceleration structures, wide BVHs, fixed-point arithmetic, and compression techniques.

\paragraph*{Ray Coherence and Stream Processing.} Early work by Wald et al.~\cite{wald_interactive_2001} demonstrated substantial performance gains through coherent ray tracing by processing packets of rays together. This paradigm was extended by Reshetov et al.~\cite{reshetov_multi-level_2005} with multi-level ray tracing, which amortizes traversal costs across multiple rays.

For incoherent rays, Wald et al.~\cite{wald_simd_2007} introduced SIMD ray
stream tracing, where rays are reorganized on-the-fly to maximize SIMD
efficiency. Boulos et al.~\cite{boulos_adaptive_2008} further refined this
approach with adaptive ray packet reordering based on runtime coherence. Stream
filtering techniques~\cite{gribble_coherent_2008,ramani_streamray_2009}
provided additional performance by processing large numbers of rays against the
same nodes, though at the cost of requiring uniform traversal orders for all
rays.
Tsakok~\cite{tsakok_faster_2009} explored multi-BVH ray stream tracing to reduce memory bandwidth while eliminating filtering costs. Later, Barringer and Akenine-Möller~\cite{barringer_dynamic_2014} presented dynamic ray stream traversal which allows rays to follow individual traversal paths while still benefiting from stream processing. Their approach demonstrated significant performance improvements by extracting implicit coherence from seemingly incoherent workloads, a key insight that our work leverages.
Fuetterling et al.~\cite{fuetterling_efficient_2015} presented efficient ray tracing kernels for modern CPU architectures that combine aspects of packet and single-ray approaches, showing that hybrid methods can outperform specialized solutions.

\paragraph*{Wide BVHs and SIMD Optimization.} Wide BVHs, which use nodes with more than two children, have been explored as a means to improve SIMD utilization and reduce tree depth. Wald et al.~\cite{wald_getting_2008} demonstrated that multi-branching BVHs enable efficient SIMD single-ray traversal. Áfra~\cite{afra_faster_2013} later extended this work with optimizations specifically for 8-wide AVX instructions.

Ernst and Greiner~\cite{ernst_2008} introduced multi bounding volume hierarchies, demonstrating how alternative BVH organizations can improve traversal efficiency through different node arrangements. Dammertz et al.~\cite{dammertz_2008} presented shallow bounding volume hierarchies specifically optimized for fast SIMD ray tracing of incoherent rays, showing that reducing tree depth while maintaining effective pruning can significantly improve performance for divergent ray workloads.

Vaidyanathan et al.~\cite{vaidyanathan_wide_2019} proposed a wide BVH traversal technique with a shortened stack to reduce memory traffic. Our approach builds upon these findings, with particular inspiration from Ylitie et al.~\cite{ylitie_efficient_2017}, who demonstrated significant memory traffic reduction through compressed wide BVHs. Their work showed that internal nodes with multiple children can be efficiently packed and processed in SIMD, while simultaneously reducing memory consumption compared to binary BVHs.

\paragraph*{Reduced Precision and Compression Techniques.} Memory-efficient acceleration structures have been explored through various compression techniques. Mahovsky and Wyvill~\cite{mahovsky_memoryconserving_2006} presented a method for hierarchically encoding BVHs using reduced precision, demonstrating that ray tracing does not always require full floating-point precision.
Mahovsky et al.~\cite{mahovsky_ray_2005} further explored reduced-precision BVHs, showing that quantization can effectively reduce memory requirements with minimal visual impact. Their approach maintained full precision for ray parameters while using quantized data structures.

Building on mesh quantization approaches, Segovia and Ernst~\cite{segovia_2010} introduced memory efficient ray tracing with hierarchical mesh quantization, demonstrating how geometry can be efficiently compressed while maintaining traversal performance. More recently, geometry compression techniques have advanced with specialized formats like DGF~\cite{barczak_2024}, which provides a dense, hardware-friendly format for lossily compressing meshlets with arbitrary topologies, enabling efficient storage and processing of complex geometry.

For subdivision surfaces and high-detail geometry, specialized compression schemes have been developed. Lier et al.~\cite{lier_2018} presented a high-resolution compression scheme for ray tracing subdivision surfaces with displacement, showing that even complex procedural geometry can be efficiently compressed. Complementing this, Benthin and Peters~\cite{benthin_2023} addressed real-time ray tracing of micro-polygon geometry with hierarchical level of detail, which adapts geometric complexity based on viewing conditions to optimize memory usage.

Fixed-point representations for ray tracing have been investigated by several researchers. Hanika and Keller~\cite{hanika_hardware_2007} demonstrated hardware ray tracing using fixed-point arithmetic, while Heinly et al.~\cite{heinly_integer_2009} explored integer-based ray tracing. More recently, Hwang et al.~\cite{hwang_mobile_2015} proposed a hybrid number representation approach for mobile ray tracing that combines aspects of fixed-point and floating-point arithmetic.

Our work integrates and extends these previous approaches in several ways: Like Ylitie et al.~\cite{ylitie_efficient_2017}, we use wide BVHs to reduce memory traffic, but we extend their compression scheme to include triangle data as well. We adopt the ray stream approach of Barringer and Akenine-Möller~\cite{barringer_dynamic_2014} to further reduce memory traffic from traversal stacks.
Unlike previous reduced-precision approaches that typically decompress data before intersection tests, we perform fixed-point ray tracing directly on the compressed structures, similar to Hanika and Keller~\cite{hanika_hardware_2007}. Our approach carefully manages precision limitations through local coordinate systems with power-of-two scaling, avoiding the geometric holes sometimes encountered with floating-point methods.
By combining these techniques---quantized wide BVHs, direct fixed-point traversal,
and ray stream processing---our method achieves greater memory traffic reduction
than previous works while maintaining high computational throughput and
numerical robustness.

The following sections detail our approach:
\Cref{sec:quantization_scheme} discusses
quantization of the bounds, the geometry, and the rays, and
\Cref{sec:fixed-point-traversal} the fixed-point ray stream traversal.

\section{Quantized BVH and Triangle Representation}


\label{sec:quantization_scheme}

Traditional BVH nodes typically store bounding box coordinates of their children as 32-bit floating-point values, consuming 24 bytes per box (2 points $\times$ 3 coordinates $\times$ 4 bytes). Similarly, triangle vertices require 36 bytes (3 vertices $\times$ 3 coordinates $\times$ 4 bytes). Our quantization scheme reduces this significantly by using 8-bit fixed-point representations within locally defined coordinate systems.

\begin{figure}[t]
  \centering
  \includegraphics[width=\linewidth]{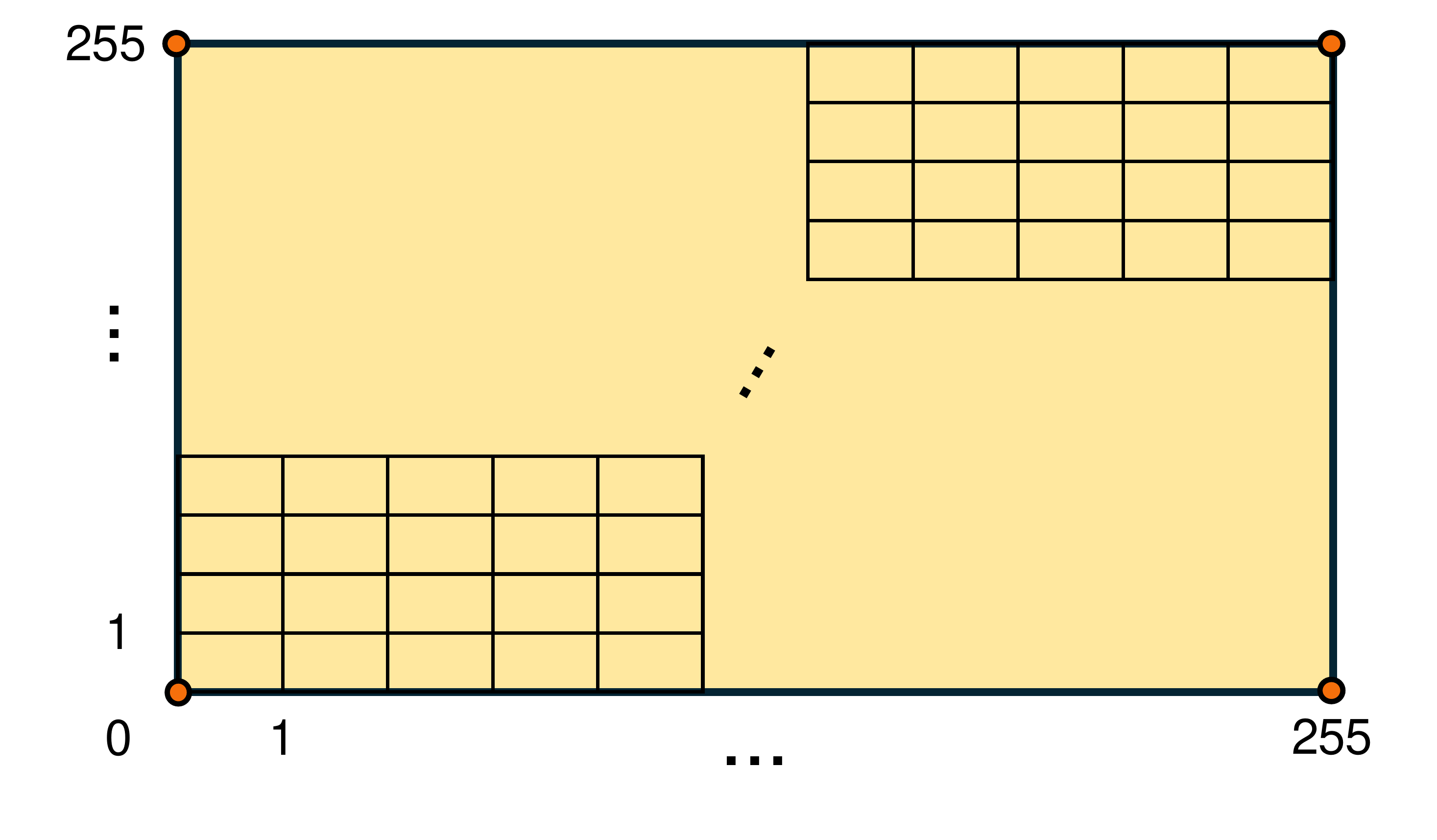}
  \caption{\label{fig:localgrid}
           Each node spans a local grid with coordinates in 8 bit. The resolution of the grid is defined by the scale factors of each axis. In this example, this results in slightly rectangular grid cells, matching the shape of the underlying bounding box. The origin (bottom left corner) of the local grid is in integer world space. Child bounds and triangles are snapped to this local grid and thus represented with 8 bits per coordinate. }
\end{figure}

\paragraph*{Local Coordinate Systems.}
For each BVH node, we define a local coordinate system with the following components:

\begin{itemize}
    \item Scale factors: Power-of-two scale for each axis (3$\times$1 byte)
    \item Origin: A full-precision integer world-space point (3$\times$4 bytes)
    \item Quantized bounds: 8-bit coordinates for each bounding plane (6$\times$1 byte per child box)
\end{itemize}
The local origin serves as the reference point for all quantized coordinates within the node. By storing this origin in full precision, we maintain positioning accuracy while allowing internal offsets to be represented with reduced precision
(see \Cref{fig:localgrid}).

\paragraph*{Scale Factors.}
For each node, we compute scale factors as powers of two, ensuring maximum
precision within the 8-bit range. This approach is equivalent
to the compression performed by Ylitie et al.~\cite{ylitie_efficient_2017} for
floating-point bounding boxes. The scale for each axis is determined by:
\begin{equation}
\label{eq:scalefactors}
scale_{axis}=\left \lceil{\log_2\frac{maxBounds_{axis} - minBounds_{axis}}{2^8 - 1}}\right \rceil.
\end{equation}


\paragraph*{Origin.}
For the root node, the origin is calculated from the lower point $\textit{p}$ of the floating-point bounding box by rounding it down to the next fixed-point number using $scale_{axis}$ bits of precision:

\begin{equation}
\label{eq:origin}
origin_{axis}=\left \lfloor{\frac{p_{axis}}{2^{scale_{axis}}}}\right \rfloor.
\end{equation}

For all other nodes, origins are provided by the parent node.

\paragraph*{Quantization of Bounds.}
Each node assigns a suitable origin and scale factors to its children. The floating-point bounds $p_{lo}$, $p_{hi}$ of the child nodes are quantized to fixed-point using the scale factor of the node as precision:

\begin{equation}
    lo_{axis} = \left \lfloor\frac{p_{lo,axis}-origin_{axis}}{2^{scale_{axis}}}\right \rfloor,
\end{equation}
\begin{equation}
    hi_{axis} = \left \lceil\frac{p_{hi,axis}-origin_{axis}}{2^{scale_{axis}}}\right \rceil.
\end{equation}

Rounding down for lower bounds and rounding up for upper bounds ensures conservatively that child nodes are contained within their parents. The resulting coordinates are guaranteed to be within $[0,255]$ due to the adaptive scale.

If a child node extent is small, the scale factors are adapted. In this case,
the child node might need less bits to represent its range.
We make sure every leaf node has the same quantization gaps and enforce 
the quantization gaps of inner nodes to be at least as large as their children.
This is done to achieve hole-free meshes, as discussed later in this section.
%


\paragraph*{Quantization of Triangles.}
For leaf nodes, we apply the same quantization principles used for child
bounding boxes. Each triangle vertex is quantized to the leaf node's local
coordinate grid using the node-specific scale factors for each axis. This
ensures that triangle geometry is represented with the same precision
as its bounds. With this scheme, each vertex coordinate
requires 8 bits of storage, resulting in a compact representation of 9
bytes per triangle (3 vertices $\times$ 3 coordinates $\times $1 byte). This represents a
significant reduction from the 36 bytes typically required in traditional ray
tracers that use full-precision floating-point coordinates. Other vertex attributes used for shading are stored seperately and are not part of the comrpession scheme. Thus, only the compact representation of a triangle is passed to the intersection tests.


Traditional vertex/index representations for quad meshes typically require around 28 bytes per quad (16 for indices, 12 for amortized vertex data) with double indirection complexity. Our quantized approach uses only 18 bytes per quad-equivalent while eliminating indirection overhead. Even optimized quad meshes with 8-byte vertices would consume more memory than this approach making it beneficial for memory-constrained environments.

This compact triangle representation enables a potential optimization: in a unified node structure, the 32 bytes used for child references in internal nodes could store up to three triangles (27 bytes) directly within leaf nodes containing three or fewer triangles. This would eliminate the need to access separate triangle storage, potentially reducing memory traffic by keeping frequently accessed small triangle groups in the node structure itself.



\paragraph*{Hole-free meshes.}
To maintain hierarchical integrity, we must ensure that child nodes never extend beyond their
parents after quantization. This is inherently guaranteed by our compression
scheme, as the scale factors of child nodes are always finer or equal to those
of their parents. However, conservative rounding of child bounding boxes can
create new overlaps between sibling nodes, potentially increasing the number of
ray-box intersections during traversal.

Our approach to maintaining the hierarchical scale factor constraint (child nodes must not use
coarser scaling than parents) is to propagate the established leaf-level scale
factors upward through the tree. Inner nodes are adjusted to use scale factors
at least as coarse as the maximum among their descendant leaf nodes. When a
node's scale factors are adjusted, we re-quantize its child bounding boxes to
match the new coordinate system.
This approach guarantees both correctness and visual fidelity while maintaining
the memory efficiency of our quantized representation.

A more subtle challenge arises with triangle quantization: For watertight
surfaces, triangles sharing an edge must remain connected after quantization. If
adjacent triangles reside in different leaf nodes with different scale factors,
they could become disconnected in fixed-point space, creating geometric holes.
To prevent this, we identify the largest scale factors for each axis among all leaf nodes and
broadcast these factors to all leaf nodes. This ensures consistent triangle
quantization across the entire scene. 
Note that this approach requires no preprocessing and guarantees correctness for all input meshes. However, scenes containing large triangles will force coarser quantization throughout the entire structure, potentially producing more visible quantization artifacts. Pre-subdividing meshes with large triangles would enable finer scaling factors across the scene, resulting in better visual quality while maintaining our memory efficiency benefits.

\begin{figure}[t]
  \centering
  \begin{tabular}{|l|l|l|}
    \hline
    \multicolumn{3}{|l|}{\textbf{struct BVHNode8}} \\
    \hline
    \multicolumn{3}{|l|}{\textbf{Children Bounds} (192 bytes total)} \\
    \hline
    float & lo\_x[8] & 32 bytes \\
    float & hi\_x[8] & 32 bytes \\
    float & lo\_y[8] & 32 bytes \\
    float & hi\_y[8] & 32 bytes \\
    float & lo\_z[8] & 32 bytes \\
    float & hi\_z[8] & 32 bytes \\
    \hline
    \multicolumn{3}{|l|}{\textbf{Union} (32 bytes total)} \\
    \hline
    \multicolumn{3}{|l|}{\emph{struct LeafNode}} \\
    \hline
    uint32\_t & primitiveOffset & 4 bytes \\
    uint32\_t & numPrimitives & 4 bytes \\
    uint32\_t & leafPad[6] & 24 bytes \\
    \hline
    \multicolumn{3}{|l|}{\emph{struct InnerNode}} \\
    \hline
    int32\_t & childOffsets[8] & 32 bytes \\
    \hline
        \textbf{NodeType} & type & 1 byte \\
    \hline
    \multicolumn{2}{|r|}{\textbf{Total Size:}} & \textbf{228 bytes} \\
    \hline
  \end{tabular}
  \caption{Memory layout of the BVHNode8 structure used in our acceleration structure. The toal size as listed is 225 bytes, which is padded to 228 bytes.}
  \label{fig:bvh8-node}
\end{figure}

\paragraph*{Memory Layout.}
\Cref{fig:bvh8-node} illustrates a typical floating-point node layout for an
8-wide BVH. In this standard representation, 192 bytes are required to store the
six bounding planes (min/max for each axis) of eight children, with each
coordinate stored as a 32-bit float. Additionally, the node stores eight 4-byte
child indices, consuming another 32 bytes, for a total of 228 bytes per node.
This layout is typically shared between internal nodes and leaf nodes, with leaf
nodes repurposing the child index space to store primitive counts and an index
to contiguously stored primitives.

\Cref{fig:bvh8-node-comp} depicts our compressed node structure. While we still
represent six bounding planes for each of the eight children, each coordinate is
stored using only 8 bits mapped to the unsigned [0,255] range. This reduces the
space required for bounds to just 48 bytes--a 75\% reduction from the
uncompressed version. The 32 bytes for child and primitive indexing remain
unchanged, as this information is directly related to the size of the scene. To
establish the local coordinate system within which the quantized bounds are
valid, we store a fixed-point origin point in 32 bit per axis (12 bytes) and
three 8-bit exponents representing power-of-two scale factors for each axis (3
bytes). The resulting compressed node requires only 96 bytes, achieving a
reduction of over 57\% compared to the uncompressed version.

Nodes for 2-wide and 4-wide BVHs follow the same organizational principle,
differing only in the number of children they contain. For the 2-wide BVH, the node size is reduced from 64 to 36 bytes, a reduction of approx. 44\%. For the 4-wide BVH, the node size is reduced from 116 to 56 bytes, a reduction of approx. 52\%. Since each node carries the fixed overhead of the quantization frame (origin and scale factors) and
indexing information, the relative space efficiency improves with higher
branching factors. The 8-wide BVH nodes achieve the most compact representation,
as they amortize this fixed overhead across more children. Nevertheless, 2-wide
and 4-wide BVH nodes still provide substantial size reductions compared to their
uncompressed counterparts.


\begin{figure}[t]
  \centering
  \begin{tabular}{|l|l|l|}
    \hline
    \multicolumn{3}{|l|}{\textbf{struct BVHNode8Comp}} \\
    \hline
    \multicolumn{3}{|l|}{\textbf{Children Bounds} (48 bytes total)} \\
    \hline
    uint8\_t & lo\_x[8] & 8 bytes \\
    uint8\_t & hi\_x[8] & 8 bytes \\
    uint8\_t & lo\_y[8] & 8 bytes \\
    uint8\_t & hi\_y[8] & 8 bytes \\
    uint8\_t & lo\_z[8] & 8 bytes \\
    uint8\_t & hi\_z[8] & 8 bytes \\
    \hline
    \multicolumn{3}{|l|}{\textbf{Union} (32 bytes total)} \\
    \hline
    \multicolumn{3}{|l|}{\emph{struct LeafNode}} \\
    \hline
    uint32\_t & primitiveOffset & 4 bytes \\
    uint32\_t & numPrimitives & 4 bytes \\
    uint32\_t & leafPad[6] & 24 bytes \\
    \hline
    \multicolumn{3}{|l|}{\emph{struct InnerNode}} \\
    \hline
    int32\_t & childOffsets[8] & 32 bytes \\
    \hline
    \multicolumn{3}{|l|}{\textbf{Quantization Data} (15 bytes total)} \\
    \hline
    int32\_t & origin[3] & 12 bytes \\
    int8\_t & e[3] & 3 bytes \\
    \hline
    \textbf{NodeType} & type & 1 byte \\
    \hline
    \multicolumn{2}{|r|}{\textbf{Total Size:}} & \textbf{96 bytes} \\
    \hline
  \end{tabular}
  \caption{Memory layout of the BVHNode8Comp structure used in our acceleration structure.}
  \label{fig:bvh8-node-comp}
\end{figure}

\paragraph*{Ray Quantization.}


%
%
%

For intersecting rays with our quantized structures, we transform each ray into
a compatible fixed-point representation. 
This is important because it reduces the memory footprint of the ray lists during ray stream tracing.
This transformation converts both the
ray origin and direction from floating-point to fixed-point format. While rays
are initially generated in floating-point representation (allowing direct
comparison with traditional methods), we apply an additional memory optimization
during conversion: the ray direction is encoded using an octahedral mapping 
technique, compressing it to a single 4-byte unsigned integer~\cite{Cigolle2014Vector}. During traversal,
this compact direction representation is decoded back into fixed-point space as
needed for intersection tests. This approach not only ensures compatibility with
our quantized acceleration structures but further reduces memory traffic
associated with ray storage.
In total, a ray consumes 32 bytes of memory: 16 bytes for an
intersection record, 12 bytes for the ray origin, and 4 bytes for the compressed direction.

\section{Fixed-Point Traversal and Intersection}
\label{sec:fixed-point-traversal}

We perform traversal and intersection operations directly on quantized data.
This section details our fixed-point algorithms for ray-box and ray-triangle
intersection.

\subsection{Fixed-Point Ray-Box Intersection}

Our ray-box intersection test operates on quantized bounds using
fixed-point arithmetic. First we move the ray and the box into the same space, using 64 bits
of precision with the same quantization gaps as the 8-bit box coordinates.
The algorithm follows the slabs method \cite{slabtest05} but adapted for
fixed-point representation.
Unlike floating-point operations where
rounding without conservative growing of the bounding boxes \cite{Ize2013BVH}
can lead to inconsistent results at boundaries (causing "cracks" or
"leaks"), our fixed-point comparison guarantees consistent outcomes. The
algorithm is outlined in \Cref{code:ray-box}.
\SetAlgoNlRelativeSize{-1} 
\SetAlCapSkip{0.5em} 
\begin{algorithm}[t]
\caption{Ray-Box Intersection}\label{code:ray-box}
\DontPrintSemicolon 
\KwIn{ray, box}
\KwOut{hit (boolean)}
$t_{min} \gets 0$\;
$t_{max} \gets MAX\_FIXED\_POINT$\;
\For{axis $\in$ \{x, y, z\}}{
    \uIf{$ray.dir[axis] = 0$}{
        \If{$ray.origin[axis] < box.min[axis]$ \textbf{or} $ray.origin[axis] > box.max[axis]$}{\tcp{Ray parallel and outside}
            \Return{false} 
        }
    }
    \Else{\tcp{Two intersections}
        $t1 \gets FixedDiv(box.min[axis] - ray.origin[axis], ray.dir[axis])$\;
        $t2 \gets FixedDiv(box.max[axis] - ray.origin[axis], ray.dir[axis])$\;
        
        \If{$ray.dir[axis] < 0$}{
            swap($t1$, $t2$)\;
        }
        
        $t_{min} \gets max(t_{min}, t1)$\;
        $t_{max} \gets min(t_{max}, t2)$\;
        
        \If{$t_{min} > t_{max}$}{\tcp{No intersection}
            \Return{false}
        }
    }
}
\Return{true}
\end{algorithm}
Special care must be taken when handling zero ray direction components in the
quantized representation. A component of the ray direction can be zero in fixed-point if the original floating-point value was too small to be represented within the
quantization precision. 
This makes this edge case appear more often than in the floating-point case and
results in directions that go parallel to one side of the bounding box.
The floating-point slab test transparently handles such cases because division by zero
results in \textit{NaN} and signed infinity which can be resolved by the same code as the
regular case. When a ray is parallel to a bounding plane, it should intersect the box if the ray passes between the minimum and maximum bounds along that axis. In fixed-point, we have to implement explicit handling for all these cases to ensure consistent ray-box testing.


\subsection{Fixed-Point Ray-Triangle Intersection}

We extend the fixed-point approach to ray-triangle intersection using an
edge-function based algorithm \cite{chirkov2005} adapted for fixed-point arithmetic.
It is based on vector triple products like many other ray triangle tests
which could have been used in this place.
\Cref{code:ray-tri} shows a simplified version of
the ray-triangle intersection algorithm.

\SetAlgoNlRelativeSize{-1} 
\SetAlCapSkip{0.5em} 

\begin{algorithm}[t]
\caption{Ray-Triangle Intersection}\label{code:ray-tri}
\DontPrintSemicolon 
\KwIn{ray, triangle, origin}
\KwOut{hit (boolean), updated ray information}
$a \gets origin + triangle.v0$\;
$b \gets origin + triangle.v1$\;
$c \gets origin + triangle.v2$\;

$ab \gets b - a$;
$ac \gets c - a$;
$bc \gets c - b$\;
$a0 \gets ray.origin - a$\;
$b0 \gets ray.origin - b$\;
$c0 \gets ray.origin - c$\;

$aN \gets Cross(ab, a0)$\; 
$bN \gets Cross(bc, b0)$\; 
$cN \gets Cross(c0, ac)$\;
$dota \gets Dot(aN, ray.direction)$\;
$dotb \gets Dot(bN, ray.direction)$\;
$dotc \gets Dot(cN, ray.direction)$\;

\If{$dota > 0$ \textbf{or} $dotb > 0$ \textbf{or} $dotc > 0$}{\tcp{Early rejection}
    \Return{false} 
}

$n \gets Cross(ab, ac)$\;
$dotn \gets Dot(ray.direction, n)$\;
$dist \gets -Dot(a0, n) / dotn$\;

\If{$dist < 0$ \textbf{or} $dist > ray.tMax$}{\tcp{Out of bounds}
    \Return{false} 
}

$ray.hitDistance \gets dist$\;
$ray.hitTriangleIndex \gets triangleIndex$\;
$ray.hitBarycentrics \gets ComputeBarycentricsFixed()$\;

\Return{true}\;
\end{algorithm}


Key challenges in our fixed-point implementation are ensuring sufficient
precision during cross product and dot product operations and avoiding rounding
before the test has made its initial decision (the dot checks).
We address these challenges by maintaining all intermediate results in full
fixed-point precision until the intersection is definitively determined. This
conservative approach guarantees that no intersections are missed along edges of
adjacent triangles, which is critical for maintaining watertight surfaces. By
postponing any precision reduction until after the intersection decision, we
prevent the numerical inconsistencies that can lead to geometric holes in
traditional floating-point implementations.

\subsection{Precision Analysis}

This section provides an analysis of the precision requirements during fixed-point ray-triangle intersection. We examine how precision requirements increase throughout the intersection test, determining the theoretical minimum bit width needed for accurate results.
Given the following range and fractional bits per component:
\begin{itemize}
    \item Ray Origin: $R_{org}$ range bits and $Q_{org}$ fractional bits
    \item Ray Direction: $R_{dir}$ range bits and $Q_{dir}$ fractional bits
    \item Triangle Vertices: $R_{tri}$ range bits and $Q_{tri}$ fractional bits
\end{itemize}
The values for the ray origin and direction are parameters and can be modified, whereas the triangle vertex precision is inferred from the BVH. In our fixed-point representation, each number uses the signed format $(R.Q)$ where $R$ bits represent the range (integer portion) and $Q$ bits represent the fractional precision. We trace precision requirements through the intersection test up to the point where the intersection is initially decided (the edge plane tests).
Fixed-point arithmetic operations affect precision as follows:
\begin{itemize}
    \item Fixed-Point Addition/Subtraction
\begin{equation}
(R_1.Q_1) \pm (R_2.Q_2): (max\{R_1, R_2\}+1).(max\{Q_1,Q_2\})
\end{equation}
    \item Fixed-Point Multiplication
\begin{equation}
(R_1.Q_1) \cdot (R_2.Q_2): (R_1+ R_2).(Q_1+Q_2)
\end{equation}
    
\end{itemize}
The rule for multiplication assumes that both operands do not take the smallest
possible negative value. In that case, one additional bit would be required.
However, this case is unlikely and avoidable by constraining the input domain or by implementing sign-aware multiplication that handles this edge case separately. Edge vectors, computed by subtracting vertices, require:
\begin{equation}
\begin{split}
    R_1&=R_{tri}+1\\
    Q_1&=Q_{tri}
\end{split}
\end{equation}
Vectors pointing from vertices to the ray origin are bounded by:
\begin{equation}
\begin{split}
    R_2&=max\{R_{tri}, R_{org}\}+1\\
    Q_2&=max\{Q_{tri},Q_{org}\}
\end{split}
\end{equation}
Normals of the edge planes computed using cross products of origin and edge vectors require (additional bits in blue):
\begin{equation}
\begin{split}
    R_3&=max\{R_{tri}, R_{org}\}+1
    \color{blue}+(R_{tri}+1)+1
    \color{black}\\
    Q_3&=max\{Q_{tri},Q_{org}\}
    \color{blue}+Q_{tri}
    \color{black}
\end{split}
\end{equation}
Finally, the dot products between edge plane normals and ray direction require (additional bits in blue):
\begin{equation}
\begin{split}
    R_4&=max\{R_{tri}, R_{org}\}+1+(R_{tri}+1)+1+
    \color{blue}R_{dir}+2
    \color{black}\\
    Q_4&=max\{Q_{tri},Q_{org}\}+Q_{tri}
    \color{blue}+Q_{dir}
\end{split}
\end{equation}
A typical scenario which we use for most of our results is
\begin{align*}
    R_{org} &= 16, &R_{dir} &= 1\\
    Q_{org} &= 8, &Q_{dir} &= 10
\end{align*}

The 8-bit vertex coordinates are added to the node origin. Since the triangle vertices are bounded by the bounding box of the leaf, we typically obtain $R_{tri} = 16$. The precision for a triangle is determined by its leaf node. In our results, $Q_{tri}=8$ is an upper bound for the leaf-level precision. Using these values, the final requirements are $R_4=38$ bits and $Q_4=26$ bits, meaning that 64 bits plus a sign bit are theoretically required to store exact intermediate results for the intersection decision. It is important that the underlying fixed-point integer data type supports sufficient bits to maintain precision throughout the calculation. In practice, the actual required precision is often substantially lower than these theoretical bounds. Geometric distributions rarely approach worst-case scenarios simultaneously in all dimensions. Given the defined scene dimensions and the fixed precision for global origins and ray directions, the exact bit-width requirements for all intermediate calculations become fully deterministic and can be precisely established prior to hardware implementation.


\subsection{Ray Stream Tracing with Wide BVHs}


Beyond the memory savings achieved through quantization, our approach further reduces memory traffic through the integration of ray stream tracing with wide BVHs. This combination minimizes traversal stack memory usage and maximizes SIMD efficiency potential, addressing key bandwidth bottlenecks in traditional ray tracing. Ray stream tracing, as introduced by Barringer and Akenine-Möller~\cite{barringer_dynamic_2014}, organizes rays into larger collections that are processed together through the acceleration structure.
Unlike traditional ray-by-ray traversal where each ray maintains its own traversal stack, ray streams consolidate traversal state across multiple rays. The key insight is that many rays, even if ultimately taking different paths through the acceleration structure, often traverse common nodes near the root. By organizing rays that need to visit the same node into groups, we can amortize node fetch costs and significantly reduce stack memory traffic.

The ray stream approach maintains a shared traversal stack containing entries that associate BVH nodes with lists of rays that need to visit them. Each stack entry contains:
\begin{itemize}
    \item a reference to a BVH node,
    \item a list of ray indices that need to process this node,
    \item traversal state information (mark children as processed).
\end{itemize}

Node data is fetched only once for potentially hundreds of rays and stack entries are shared
among rays.
In combination with our
compressed node structure, the memory traffic is further reduced. On the other hand, during ray
stream traversal, ray lists have to be accessed, too.
We thus provide a ray encoding scheme using octahedral maps in 
\Cref{sec:quantization_scheme}, reducing memory loads 
when applied within the ray stream tracing technique.

\section{Results}\label{sec:results}

Our evaluation focuses on the primary
goal of this work: minimizing memory traffic while maintaining visual fidelity.
We instrumented our code to measure memory transactions when loading bounds, geometry, and ray streams during ray tracing. 
We also provide visual evaluation of the effect of quantization on the images.

We construct our acceleration structures using a two-phase approach. Initially,
we build floating-point BVHs of widths 2, 4, and 8 using Embree~\cite{embree_2014}. Once the conventional BVH is constructed, we convert it to
our compressed representation in a post-process. This involves two iterations over the nodes.  For each node, we compute the
optimal local coordinate system as described in
\Cref{sec:quantization_scheme} in a first pass. This includes defining the origin point and scale factors that
provide maximum precision within the 8-bit quantization range. In the same step, we
quantize all bounding boxes and triangle vertices using these node-specific
local coordinate systems. The resulting quantized data is then stored in our
compact memory layout, ready for traversal using the fixed-point algorithms
detailed in \Cref{sec:fixed-point-traversal}. In the second iteration, the leaf-level precisions are propagated up the tree, ensuring a consistent hierarchy (see \Cref{sec:quantization_scheme}.

\begin{figure*}
    \centering
    \setlength{\tabcolsep}{1pt} 
    \renewcommand{\arraystretch}{1} 
    \begin{tabular}{@{}cccccc@{}} 
    
    \includegraphics[width=0.15\textwidth]{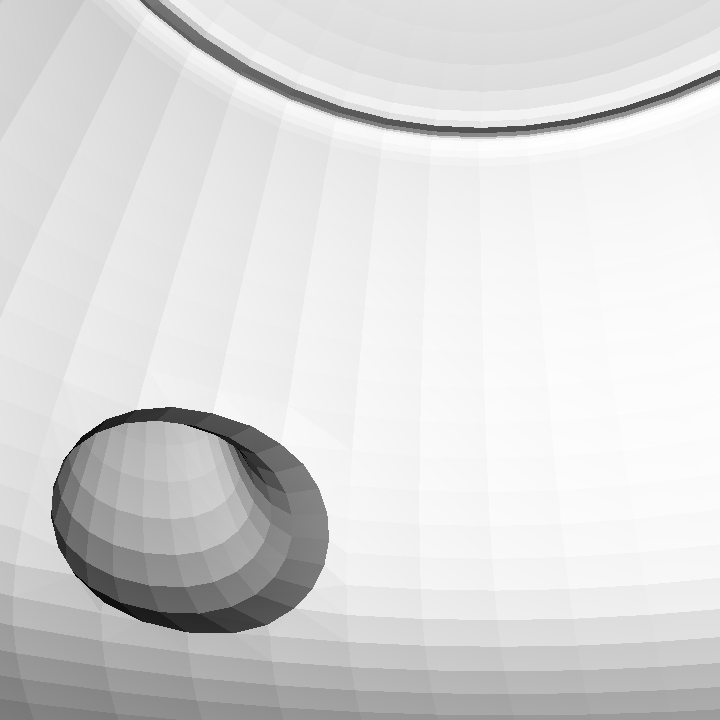} & 
    \includegraphics[width=0.15\textwidth]{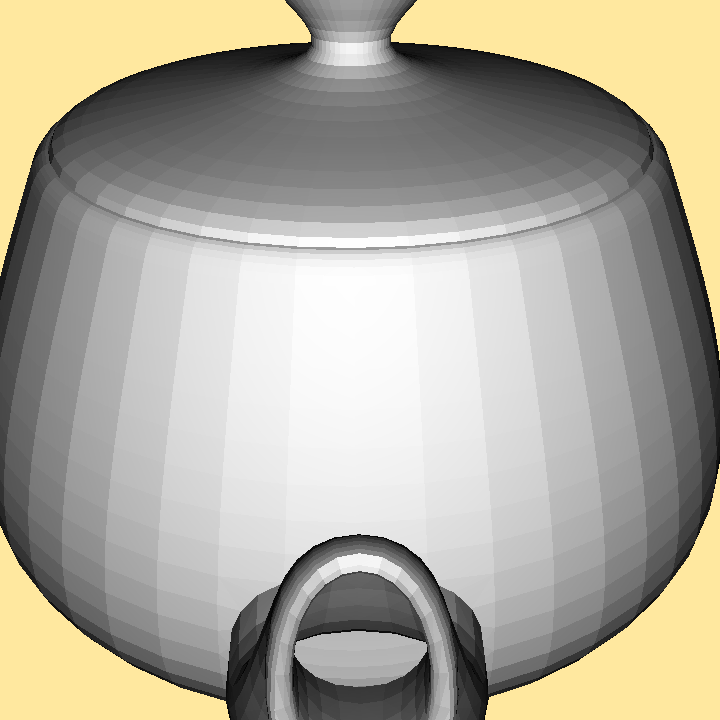} & 
    \includegraphics[width=0.15\textwidth]{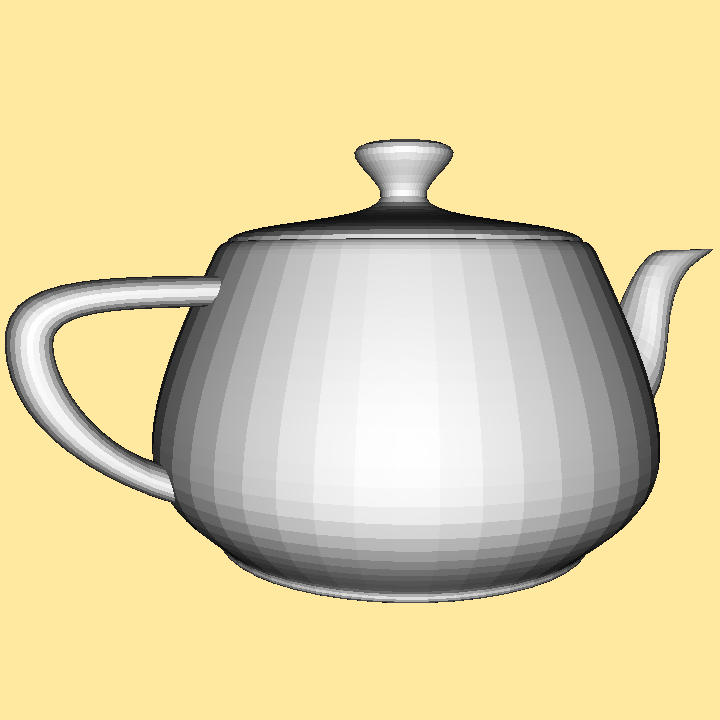} & 
    \includegraphics[width=0.15\textwidth]{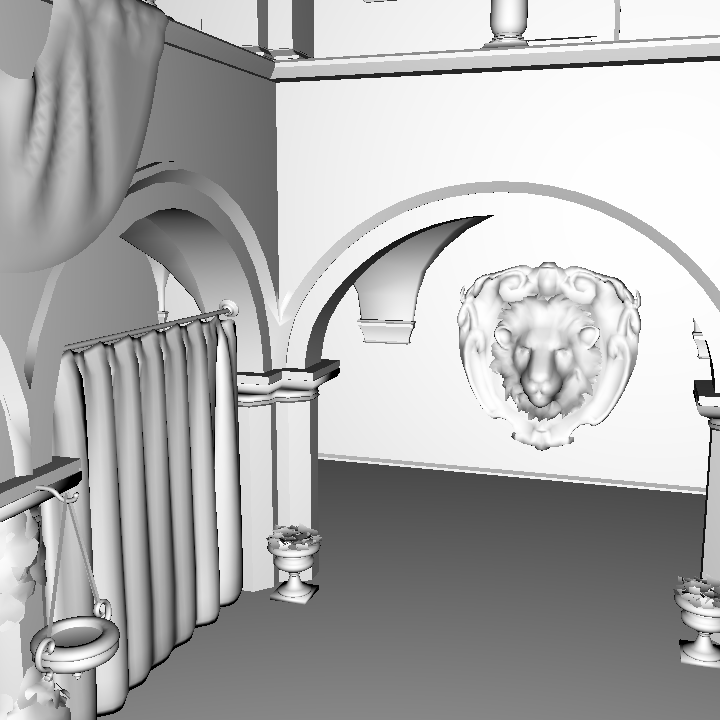} & 
    \includegraphics[width=0.15\textwidth]{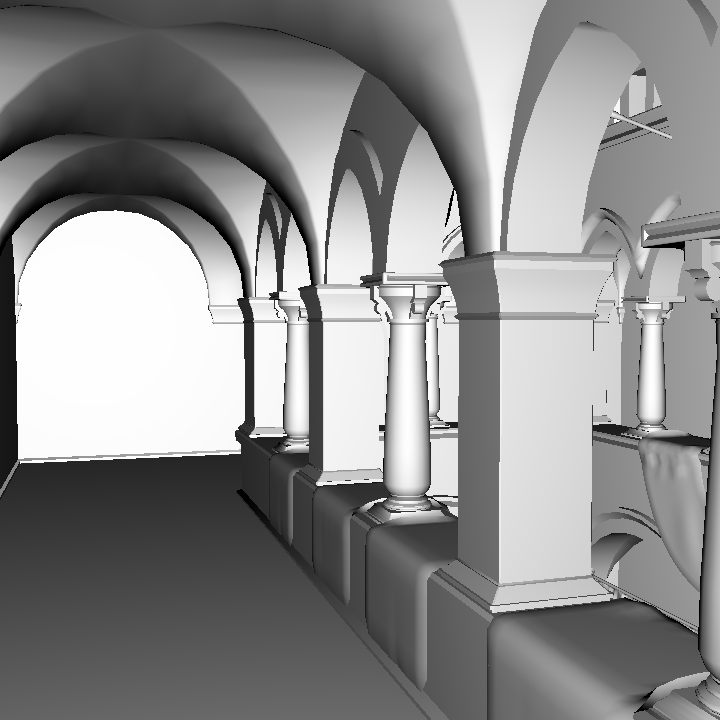} & 
    \includegraphics[width=0.15\textwidth]{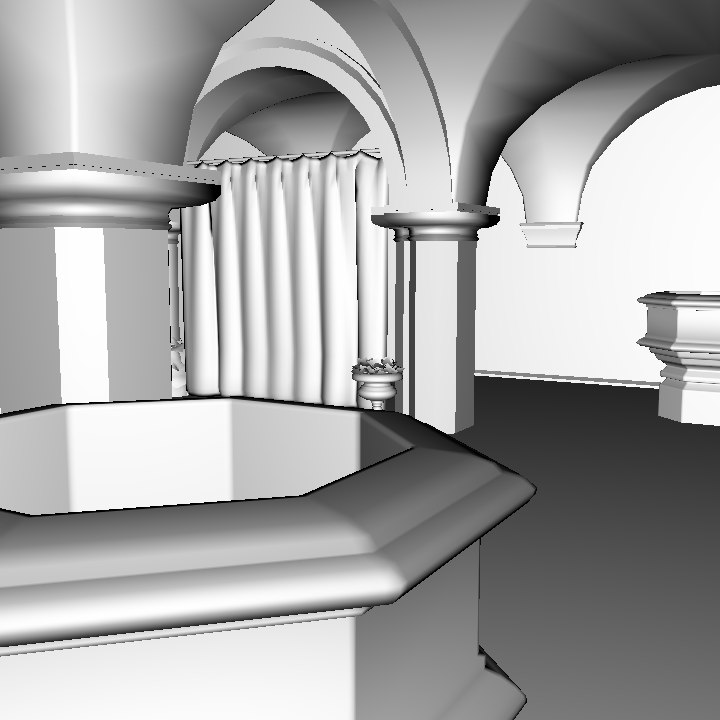} \\
    
    \scriptsize Teapot, View 1 & 
    \scriptsize Teapot, View 2 & 
    \scriptsize Teapot, View 3 & 
    \scriptsize Sponza, View 1 & 
    \scriptsize Sponza, View 2 & 
    \scriptsize Sponza, View 3 \\[2pt]
    
    \includegraphics[width=0.15\textwidth]{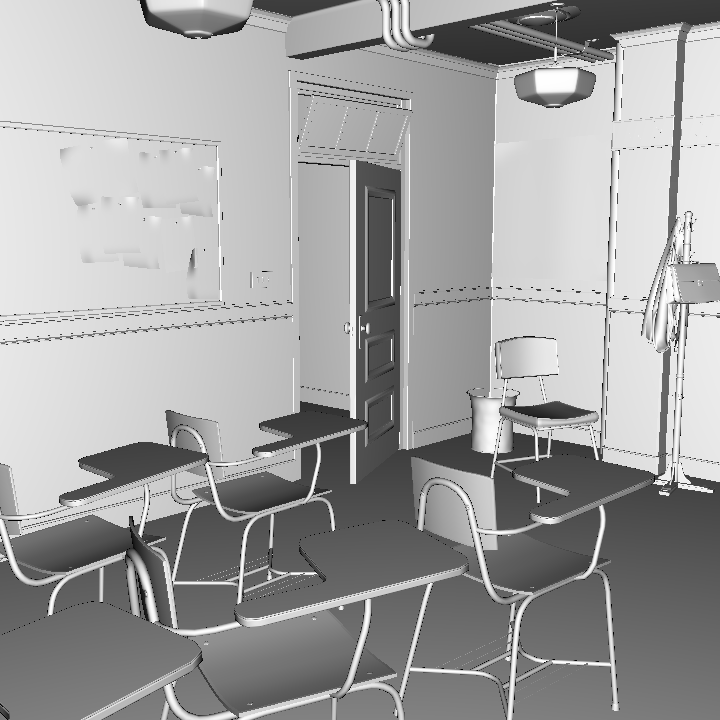} & 
    \includegraphics[width=0.15\textwidth]{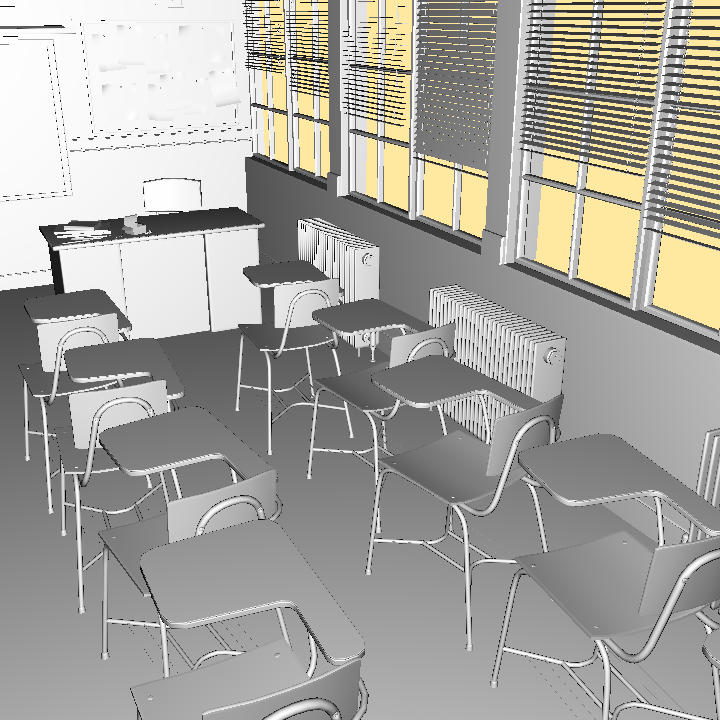} & 
    \includegraphics[width=0.15\textwidth]{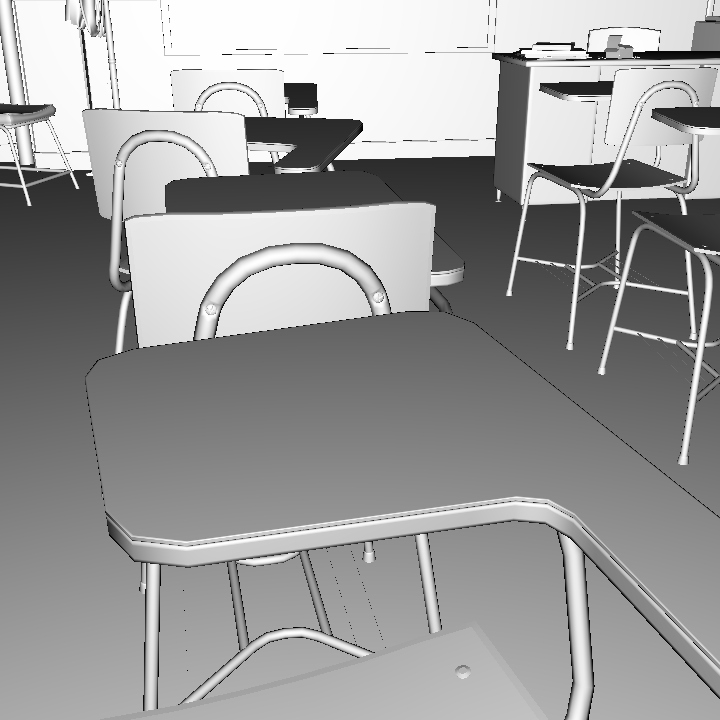} & 
    \includegraphics[width=0.15\textwidth]{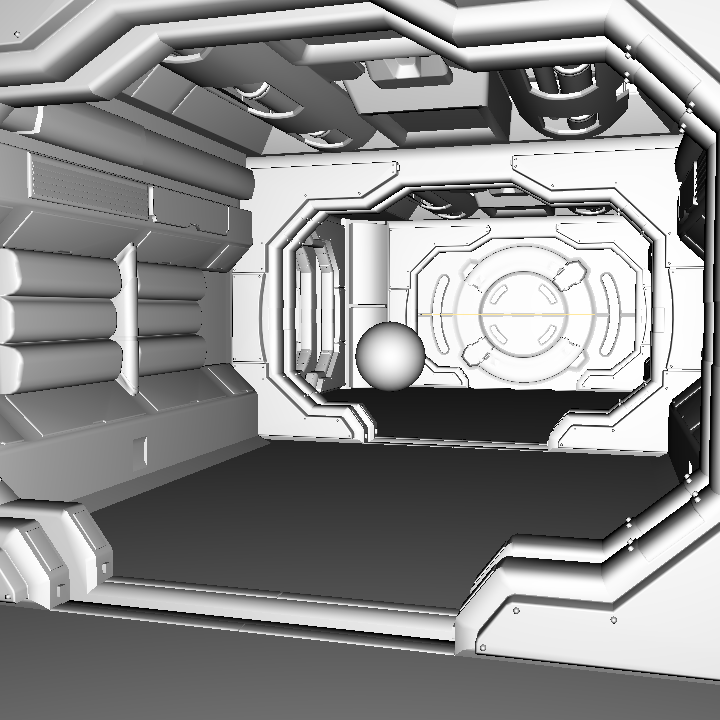} & 
    \includegraphics[width=0.15\textwidth]{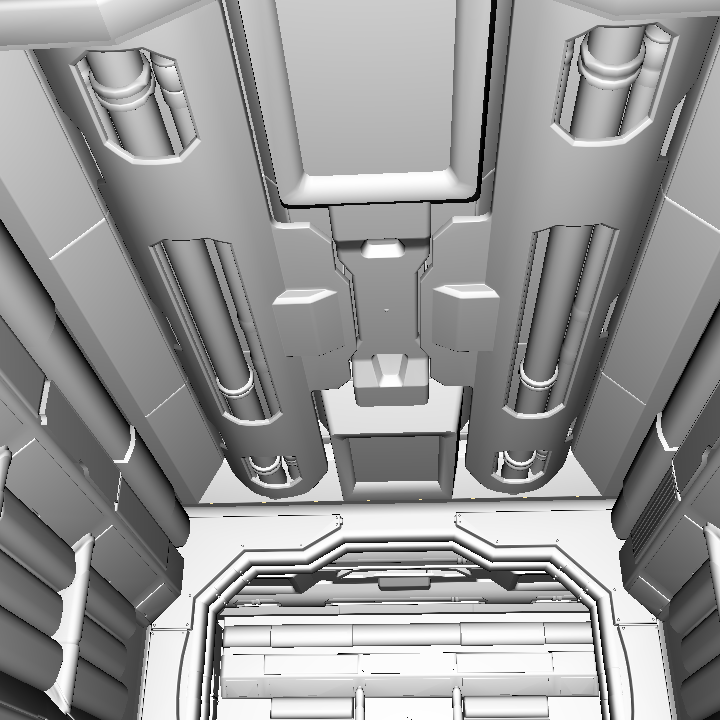} & 
    \includegraphics[width=0.15\textwidth]{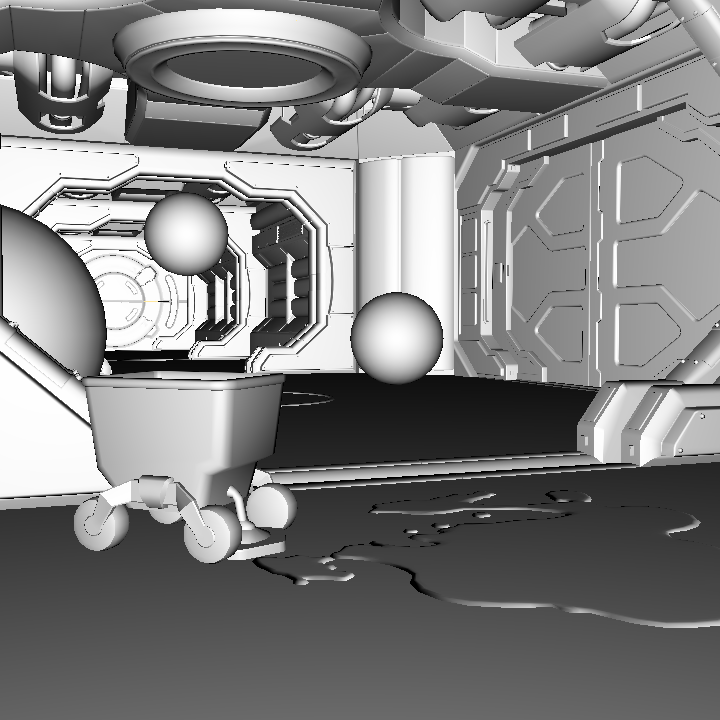} \\

    \scriptsize Classroom, View 1 & 
    \scriptsize Classroom, View 2 & 
    \scriptsize Classroom, View 3 & 
    \scriptsize Corridor, View 1 & 
    \scriptsize Corridor, View 2 & 
    \scriptsize Corridor, View 3 \\[2pt]
    
    \includegraphics[width=0.15\textwidth]{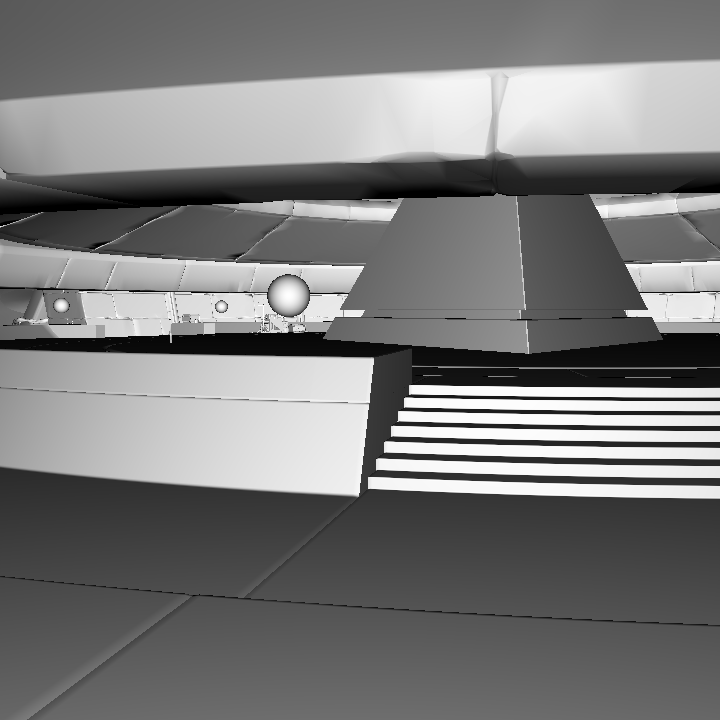} & 
    \includegraphics[width=0.15\textwidth]{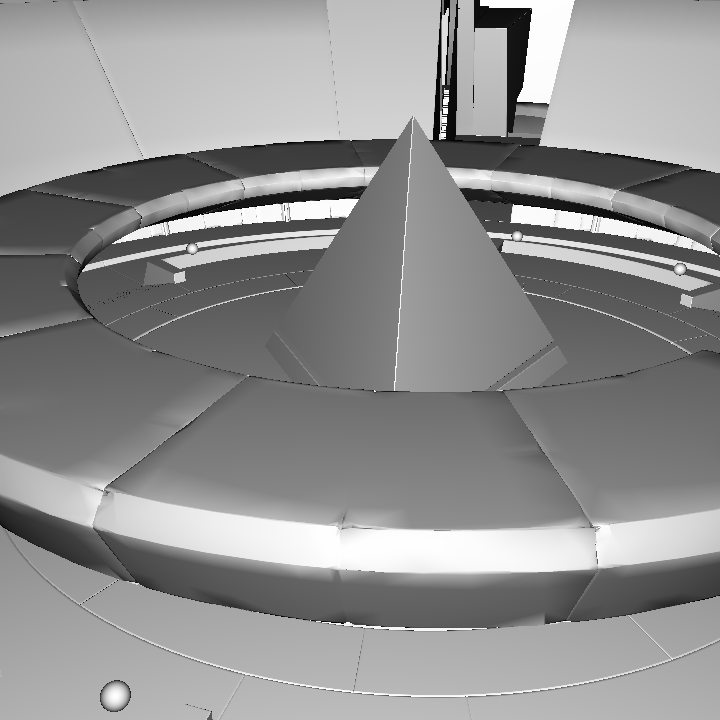} & 
    \includegraphics[width=0.15\textwidth]{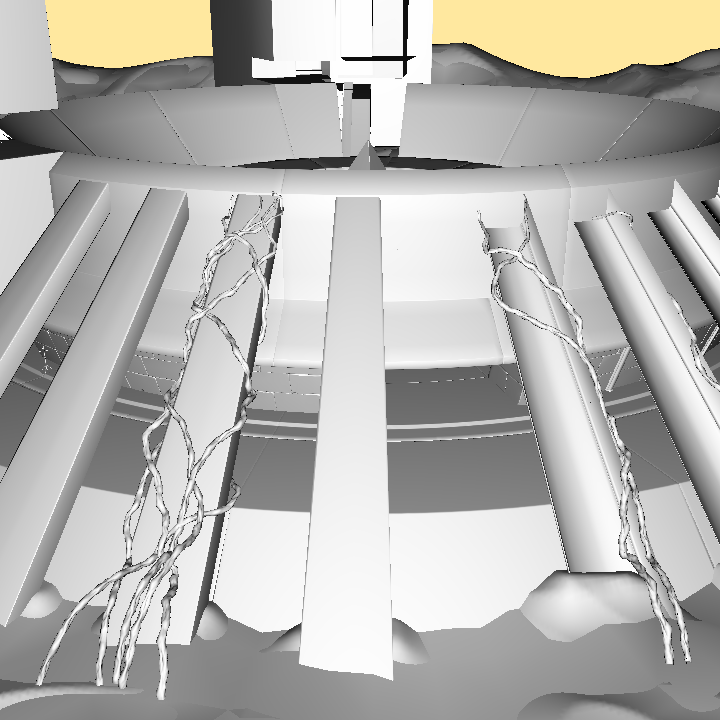} & 
    \includegraphics[width=0.15\textwidth]{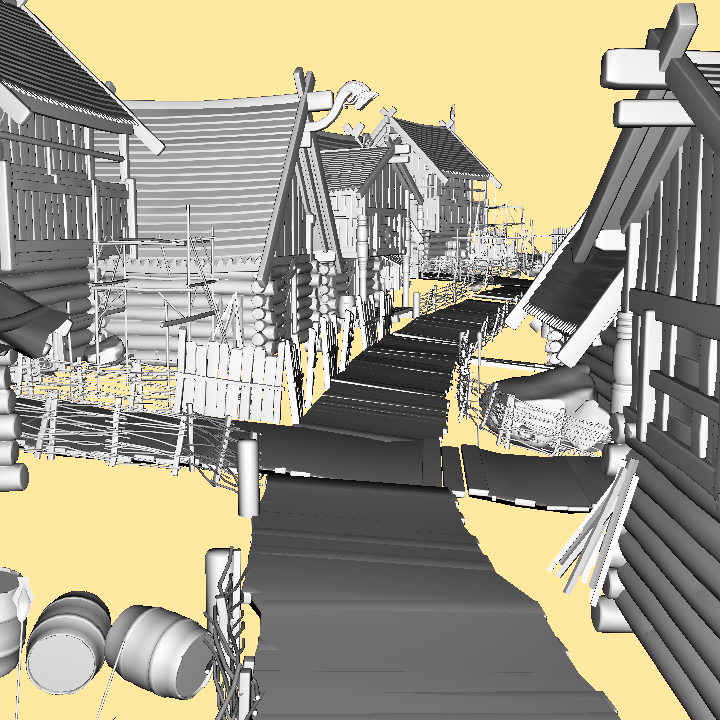} & 
    \includegraphics[width=0.15\textwidth]{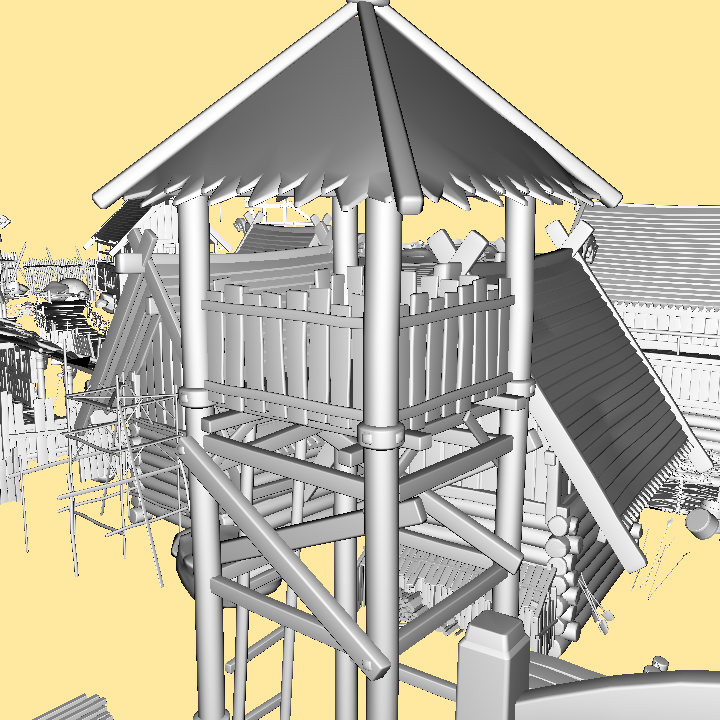} & 
    \includegraphics[width=0.15\textwidth]{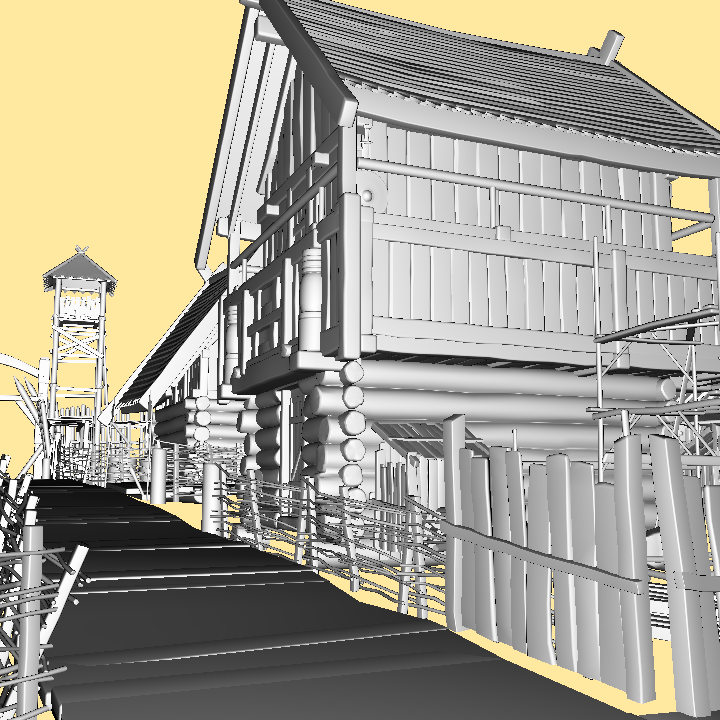} \\
    
    \scriptsize Courtyard, View 1 & 
    \scriptsize Courtyard, View 2 & 
    \scriptsize Courtyard, View 3 & 
    \scriptsize Viking, View 1 & 
    \scriptsize Viking, View 2 & 
    \scriptsize Viking, View 3 \\
    
    \end{tabular}
\caption{The test scenes used for our evaluation.
For triangle counts and BVH sizes see \Cref{tab:memory-reduction}.
\label{fig:scenes}}
\end{figure*}

\paragraph*{Test Scenes.}
For our evaluation, we used a diverse set of 6 scenes spanning different
complexity levels and characteristics. These include single object models with
moderate triangle counts, detailed indoor/room scenes with varied geometric
features, and expansive world scenes containing millions of triangles. The test
scenes are shown in \Cref{fig:scenes}.
For each scene, we list the total
node count and triangle count in \Cref{tab:memory-reduction} to provide context for the complexity.





\paragraph*{Configurations}
We conducted comparisons across 2-wide, 4-wide, and 8-wide BVH configurations.
Each type of BVH is evaluated with and without applied quantization.
All configurations are tested with both
traditional single-ray traversal and the ray stream approach.
The 12 different configurations are a combination of
\begin{itemize}
\item BVH2, BVH4, BVH8 for the number of children of the nodes,
\item SR or RS for single ray or ray stream,
\item and C for compressed and U for uncompressed.
\end{itemize}


\definecolor{cfirst}{rgb}{0.55, 0.71, 0.0} 
\definecolor{csecond}{rgb}{0.93, 0.57, 0.13}
\definecolor{cthird}{rgb}{0.77, 0.12, 0.23}
\newcommand{\marki}[1]{{\color{cfirst}\textbf{#1}}}
\newcommand{\markii}[1]{{\color{csecond}\textbf{#1}}}
\newcommand{\markiii}[1]{{\color{cthird}\textbf{#1}}}

\begin{table*}[htbp]
\centering
\begin{tabular}{|l|c|c|c|c|c|c|}
\hline
& \textbf{Teapot} & \textbf{Sponza} & \textbf{Classroom} & \textbf{Corridor} & \textbf{Courtyard} & \textbf{Viking} \\
\hline
\#Triangles             & 16k     & 262k      & 607k     & 292k     & 2.8m     & 3.9m    \\
\#BVH2 nodes            & 13.9k     & 263.4k      & 577.3k     & 290.6k     & 2.69m     & 3.65m     \\
\#BVH4 nodes            & 4.7k      & 86.3k       & 200.9k     & 101.3k     & 965.1k      & 1.33m     \\
\#BVH8 nodes            & 4.7k      & 81.7k       & 191.5k     & 94k      & 908.4k      & 1.27m     \\
\hline
Triangles uncomp. [MiB]   & 0.54    & 9.00     & 20.86   & 10.04   & 95.95   & 135.06   \\
Triangles comp. [MiB]    & 0.13    & 2.25     & 5.21    & 2.51    & 23.99    & 33.76    \\
BVH2 nodes uncomp. [MiB]  & 0.85    & 16.07    & 35.23   & 17.73   & 164.46   & 222.58   \\
BVH2 nodes comp. [MiB]   & 0.48    & 9.04     & 19.82   & 9.97   & 92.51    & 125.20   \\
BVH4 nodes uncomp. [MiB]  & 0.52    & 9.56     & 22.22   & 11.20   & 106.77   & 147.01   \\
BVH4 nodes comp. [MiB]    & 0.25    & 4.61     & 10.73   & 5.41    & 51.54    & 70.97    \\
BVH8 nodes uncomp. [MiB]  & 1.02    & 17.75    & 41.63   & 20.44   & 197.52   & 277.00   \\
BVH8 nodes comp. [MiB]   & 0.43    & 7.47     & 17.53   & 8.60    & 83.16    & 116.62   \\
\hline
\end{tabular}
\caption{An overview of scene complexity and memory savings due to compression.
Note that our BVHs are
constructed by Embree. Their focus might have been on ray tracing performance
instead of on memory footprint, so our BVH8 number of nodes is about the same as
for BVH4, resulting in a much larger memory footprint.
Ylitie et al.~\cite[Table 6]{ylitie_efficient_2017} report similar BVH4 cost for Sponza, but only 2.3~MiB for the corresponding BVH8.
\label{tab:memory-reduction}}
\end{table*}

\begin{table*}[htbp]
\centering
\begin{tabular}{|c|cc|cc|cc|cc|cc|cc|}
\hline
& \multicolumn{2}{c|}{\textbf{Teapot}} & \multicolumn{2}{c|}{\textbf{Sponza}} & \multicolumn{2}{c|}{\textbf{Classroom}} & \multicolumn{2}{c|}{\textbf{Corridor}} & \multicolumn{2}{c|}{\textbf{Courtyard}} & \multicolumn{2}{c|}{\textbf{Viking}} \\
\cline{1-13}
Configuration & Box & Tri& Box& Tri& Box & Tri& Box& Tri & Box & Tri & Box & Tri \\
\hline
BVH2-RS-C & 25         & \marki{4} & 69 & 10 & 60 & 15 & 54 & 4 & 79 & 23 & 53 & 10 \\
BVH2-RS-U & 24         & \marki{4} & 59 & \marki{5} & 46 & \marki{4} & 51 & \marki{3} & 70 & \marki{18} & 49 & \marki{8} \\
BVH2-SR-C & 25         & \marki{4} & 69 & 10 & 60 & 15 & 54 & 4 & 79 & 23 & 53 & 10 \\
BVH2-SR-U & 24         & \marki{4} & 59 & \marki{5} & 46 & \marki{4} & 51 & \marki{3} & 70 & \marki{18} & 49 & \marki{8} \\
BVH4-RS-C & \marki{23} & 7 & 63 & 16 & 53 & 19 & 50 & 9 & 84 & 31 & 53 & 14 \\
BVH4-RS-U & \marki{23} & 7 & 55 & 9 & \marki{43} & 6 & 47 & 7 & 76 & 21 & 50 & 11 \\
BVH4-SR-C & \marki{23} & 7 & 63 & 15 & 52 & 18 & \marki{49} & 8 & 78 & 27 & 51 & 13 \\
BVH4-SR-U & \marki{23} & 7 & 57 & 10 & \marki{43} & 5 & 47 & 6 & 71 & 19 & \marki{48} & 10 \\
BVH8-RS-C & 40         & 9 & 87 & 22 & 74 & 25 & 74 & 11 & 148 & 48 & 93 & 22 \\
BVH8-RS-U & 38         & 8 & 75 & 11 & 56 & 8 & 71 & 8 & 132 & 33 & 92 & 18 \\
BVH8-SR-C & 35         & 6 & 83 & 16 & 66 & 19 & 68 & 8 & 112 & 28 & 75 & 13 \\
BVH8-SR-U & 34         & 6 & \marki{55} & 9 & 52 & 6 & 65 & 6 & \marki{70} & \marki{18} & 70 & 10 \\
\hline

\end{tabular}
\caption{Box and triangle intersection counters (in millions) for 12
configurations on 6 scenes (averaged over 3 viewpoints). The BVH8-RS-C configuration produces a larger amount of box
and triangle intersections but stays relatively close to the best technique in terms of total memory traffic (\Cref{tab:total-traffic}).
\label{tab:intersection-counters}}
\end{table*}

\begin{table*}[htbp]
\centering
\begin{tabular}{|c|c|c|c|c|c|c|}
\hline
& \textbf{Teapot} & \textbf{Sponza} & \textbf{Classroom} & \textbf{Corridor} & \textbf{Courtyard} & \textbf{Viking} \\
\hline



BVH2-RS-C & 427 (86\%) & 1184 (87\%) & 1075 (82\%) & 883 (89\%) & 1411 (81\%) & 895 (83\%) \\
BVH2-RS-U & 527 (75\%) & 1211 (82\%) & 929 (81\%) & 1030 (85\%) & 1515 (68\%) & 1022 (74\%) \\
BVH4-RS-C & 219 (74\%) & 600 (76\%) & 536 (71\%) & 438 (79\%) & \marki{876} (71\%) & \marki{507} (73\%) \\
BVH4-RS-U & 267 (51\%) & 617 (64\%) & 468 (66\%) & 507 (65\%) & 932 (53\%) & 573 (56\%) \\
BVH8-RS-C & \marki{216} (70\%) & 503 (69\%) & 474 (64\%) & \marki{366} (75\%) & 947 (65\%) & 522 (67\%) \\
BVH8-RS-U & 257 (46\%) & \marki{468} (52\%) & \marki{348} (52\%) & 413 (57\%) & 955 (44\%) & 613 (45\%) \\

\hline
\end{tabular}
\caption{Measured ray access traffic (in MiB) and fraction of of total traffic (in \%) for the configurations using ray
stream tracing on 6 scenes (averaged over 3 viewpoints). The measured ray access
traffic is constant for single ray tracing since there is no ray list to manage,
rays are processed one by one. Reducing the ray size is an effective way of reducing overall traffic for the ray stream variants.
\label{tab:ray-traffic}}
\end{table*}

\begin{table*}[htbp]
\centering
\begin{tabular}{|c|c|c|c|c|c|c|}
\hline
& \textbf{Teapot} & \textbf{Sponza} & \textbf{Classroom} & \textbf{Corridor} & \textbf{Courtyard} & \textbf{Viking} \\
\hline
BVH2-RS-C & \markiii{497} & 1364         & 1308         & 993          & \markiii{1739}& 1074 \\
BVH2-RS-U & 705           & 1469         & 1141         & 1206         & 2237          & 1381 \\
BVH2-SR-C & 555           & 1514         & 1424         & 1096         & 1890          & 1161 \\
BVH2-SR-U & 1027          & 2191         & 1682         & 1816         & 3120          & 1969 \\
BVH4-RS-C & \marki{297}   & \markii{785} & 756          & \markii{551} & \marki{1229}  & \marki{693} \\
BVH4-RS-U & 525           & 969          & \markii{710} & 775          & 1749          & \markiii{1024} \\
BVH4-SR-C & 506           & 1341         & 1218         & 946          & 1852          & 1093 \\
BVH4-SR-U & 1071          & 2346         & 1617         & 1782         & 3335          & 2069 \\
BVH8-RS-C & \markii{308}  & \marki{731}  & \markiii{742}& \marki{489}  & \markii{1458} & \markii{777} \\
BVH8-RS-U & 560           & \markiii{902}& \marki{667}  & \markiii{728}& 2195          & 1352 \\
BVH8-SR-C & 657           & 1591         & 1477         & 1119         & 2383          & 1392 \\
BVH8-SR-U & 1530          & 2905         & 1995         & 2354         & 4515          & 2894 \\
\hline
\end{tabular}
\caption{Measured total traffic (ray lists, ray traversal stacks, bounds, triangles, in MiB)
for 12 configurations on 6 scenes (averaged over 3 viewpoints).
\marki{Best}, \markii{second best}, and \markiii{third best} numbers for each scene are color coded.
BVH4-RS-C and BVH8-RS-C are close contenders for the best technique, the BVH4 option being a bit better
for larger scenes.
\label{tab:total-traffic}}
\end{table*}

\paragraph*{Memory Reduction Analysis.}
An overview of the complexity of our test scenes is provided in
\Cref{tab:memory-reduction}, which also illustrates the memory savings achieved
by using the compressed triangles and BVH nodes presented in
\Cref{sec:quantization_scheme}. As expected, larger scenes experience more
significant reductions in absolute memory footprint. The largest scene in our
test suite, \textit{Viking}, contains 3.9 million triangles, requiring
approximately 135 MiB of memory for vertex coordinates alone. Our compression
technique reduces this to less than 34 MiB. This scene also demonstrates the highest absolute reduction in BVH
node size with an 8-wide BVH. The 1.27 million 8-wide BVH nodes originally
occupy 277 MiB of storage, while our compressed representation requires only
about 117 MiB, saving 160 MiB.

\Cref{tab:memory-reduction} further reveals that the node count decreases significantly when transitioning from a 2-wide to a 4-wide BVH structure. However, the reduction is much less pronounced when moving from 4-wide to 8-wide configurations. This can be attributed to two opposing factors: while the number of internal nodes decreases, there is a corresponding increase in leaf nodes and empty nodes. For instance, in the Viking scene, the Embree builder created 615,050 empty nodes. This suggests that careful optimization of BVH builder parameters could yield more suitable hierarchies that maximize the benefits of our compression scheme.

\paragraph*{Memory Traffic Analysis.}
\Cref{tab:intersection-counters} provides intersection statistics for all 
configurations, showing the number of ray-box and ray-triangle intersection
tests performed across the various scenes. These
intersection counters offer insight into the computational efficiency of each
approach, with lower values indicating reduced traversal work.

\Cref{tab:ray-traffic} shows memory access required for fetching rays during ray stream tracing. This data shows that reducing the size of rays in memory using techniques like encoding with octahedral maps is important for the ray stream variants.
\Cref{tab:total-traffic} presents a detailed breakdown of memory traffic for the
same configurations, measured in MiB transferred per frame. For single-ray
traversal, we track node accesses, ray accesses, and stack accesses separately.
For ray stream tracing, we monitor node accesses, ray accesses, ray stream stack
accesses, and ray list accesses.
All variants also include geometry access. Throughout these experiments, the configurations BVH4-RS-C and BVH8-RS-C come out as the two 
best approaches overall, with a slight preference for BVH4 for high triangle counts.

\Cref{fig:traffic-bounces} shows total memory traffic accumulated for diffuse bounces using path tracing for the presented configurations. The SR-U configurations accumulate the highest amounts of memory traffic among all candidates. The configurations BVH4-RS-C and BVH8-RS-C consistently accumulate the least memory traffic showing the effectiveness of both ray stream traversal and compression in unison. The memory traffic in configuration BVH8-RS-C amounts to only 18\% of the memory traffic produced in the respective uncompressed, single ray version BVH8-SR-U. 

\subsection{Visual Quality}

\begin{figure*}
  \centering
  \includegraphics[width=.8\linewidth]{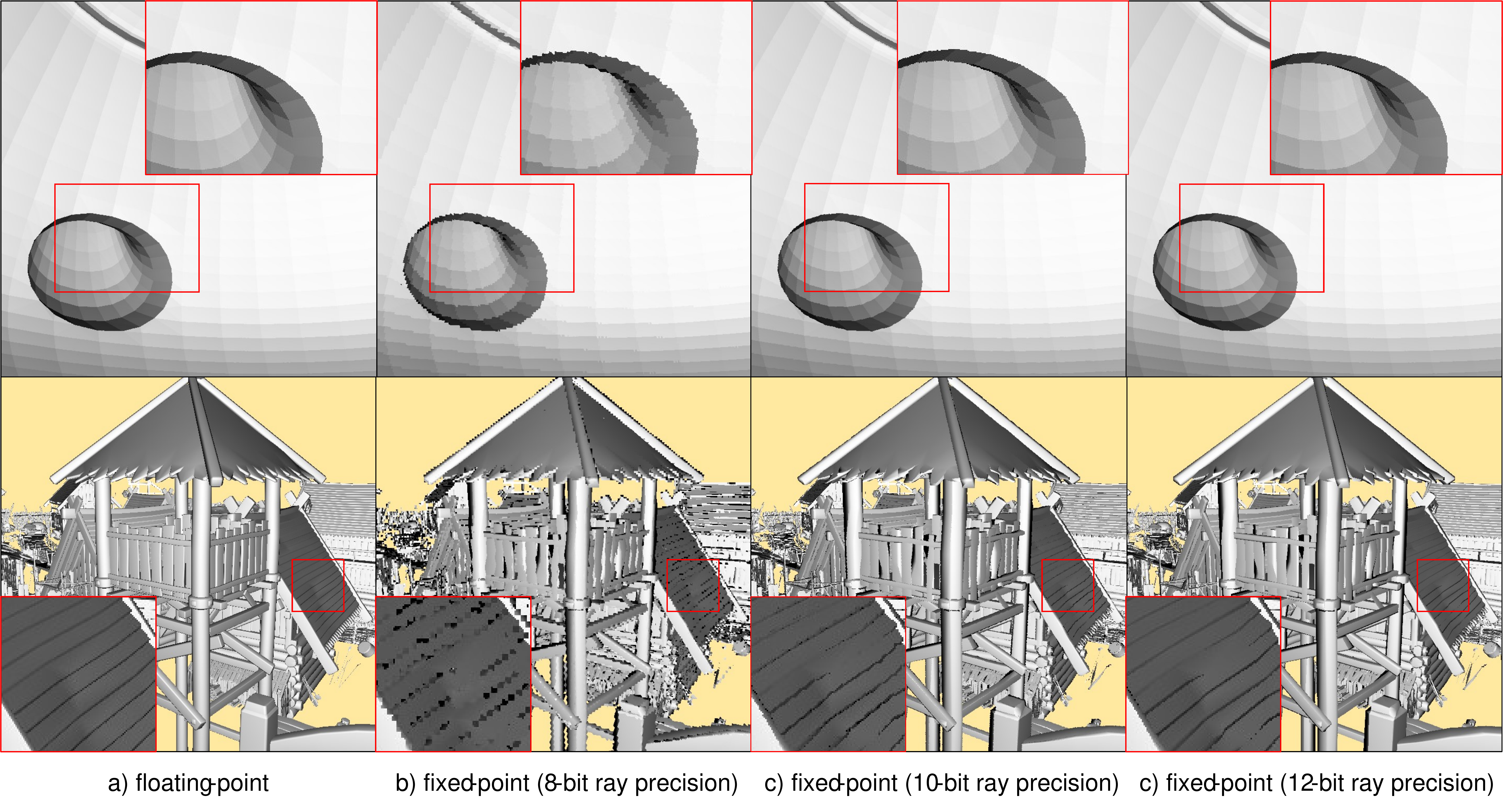}
  \caption{\label{fig:visual-quality-bits} Comparison of ray-traced images (\textit{Teapot} and \textit{Viking}) using the
  compressed BVH and triangle representations against a floating-point
  reference. Quantization of ray directions results in visible differences,
  less pronounced when using 10-bit or 12-bit ray directions. In the fixed-point images, the node origins use 32 bits
  per axis for the origins (see \Cref{fig:bvh8-node-comp}). Triangles
  and child bounding boxes use 8-bit addressing within the local frame,
  which has much less impact on the image.
            }
\end{figure*}

\begin{figure}
  \centering
  \includegraphics[width=\linewidth]{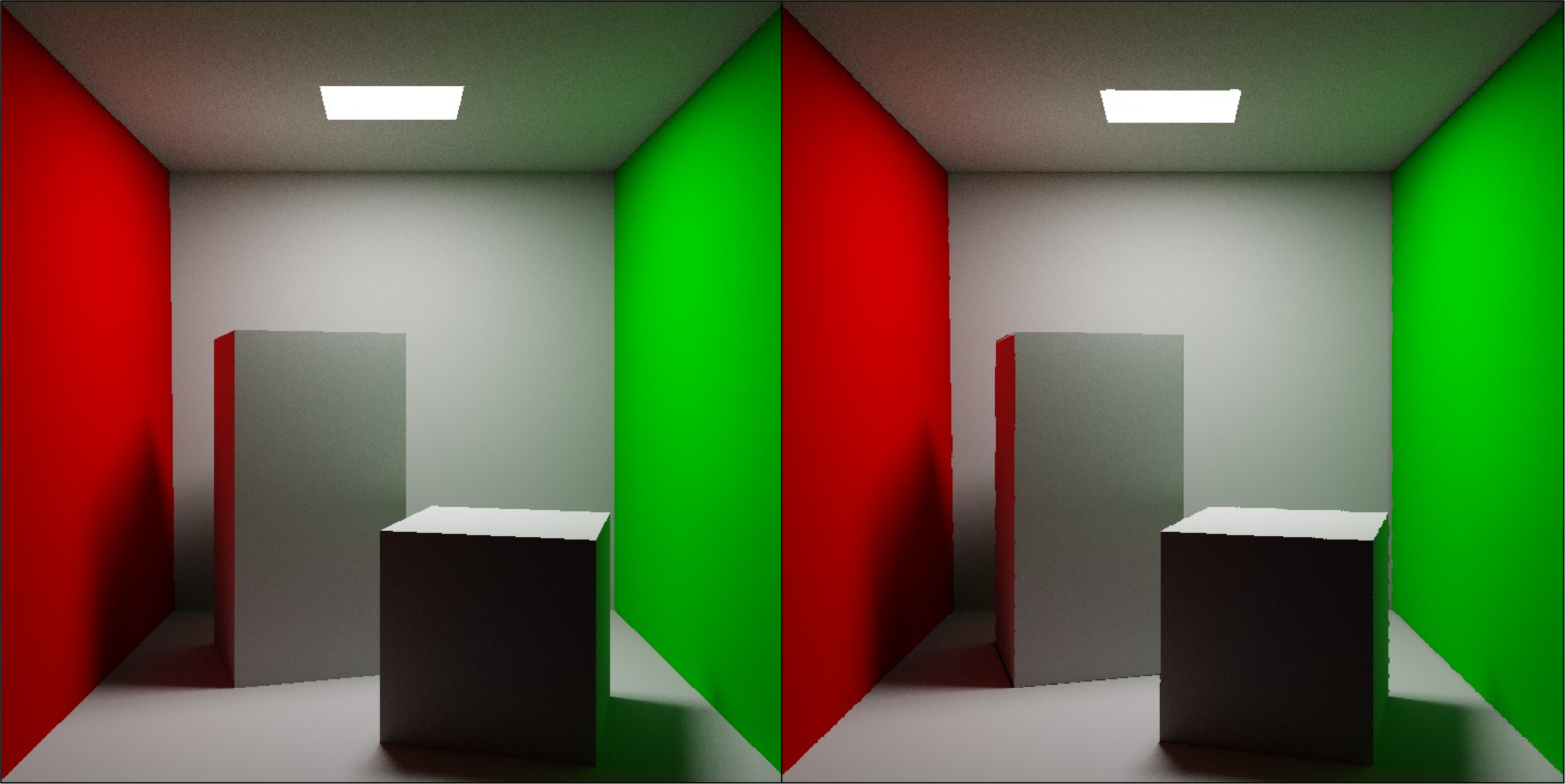}
  \caption{\label{fig:visual-quality-pt}
           Left: Path traced without compression. Right: Path traced with compression of Rays, BVH and triangles. 10-bits were used for the precision of ray origins and directions. Quantization errors can be spotted mostly along edges.}
\end{figure}

\begin{figure*}
  \centering
  \includegraphics[width=.8\linewidth]{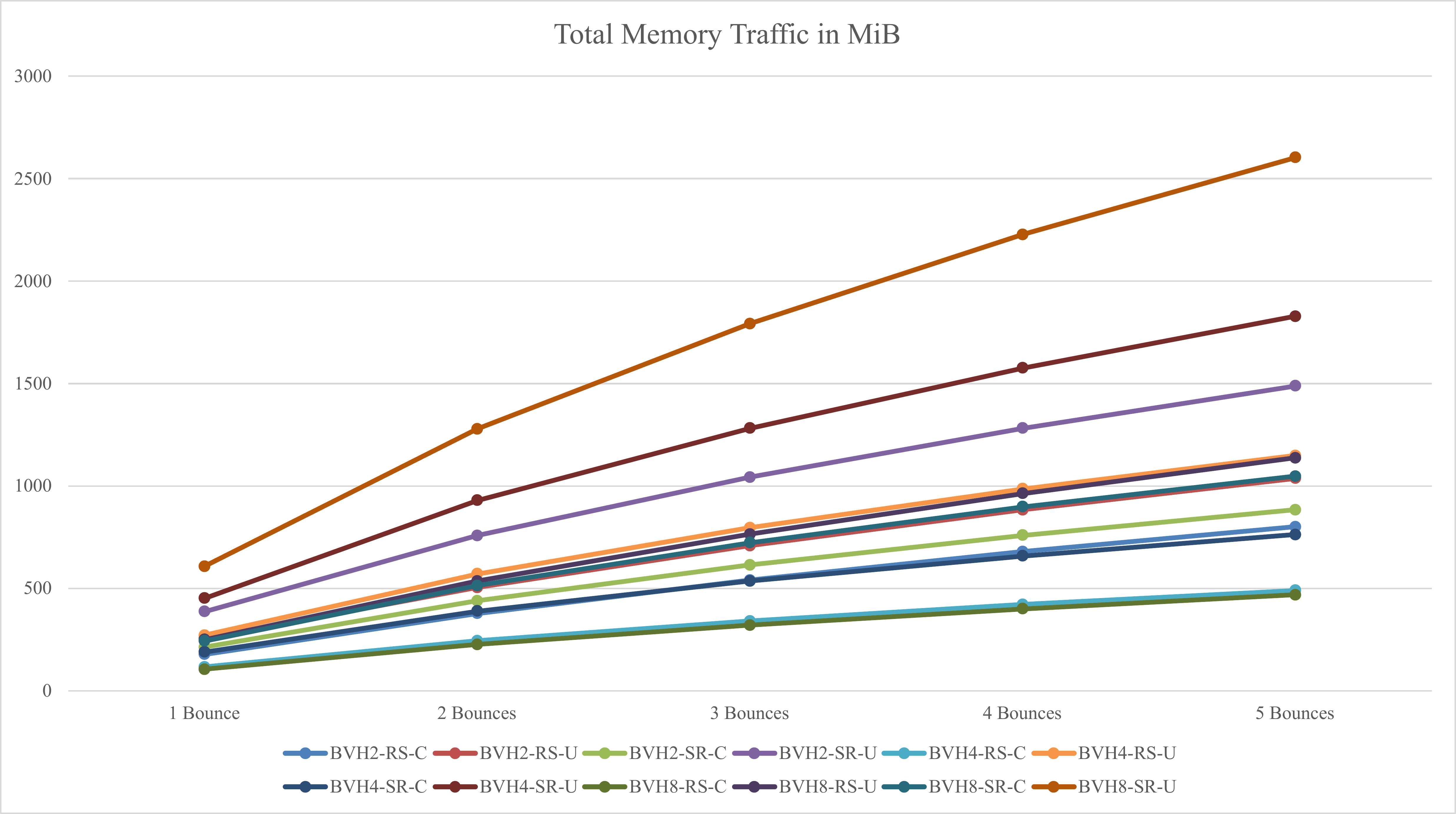}
  \caption{\label{fig:traffic-bounces}
           Comparison of the accumulated total memory traffic over multiple diffuse bounces using path tracing for the cornell box (\Cref{fig:visual-quality-pt}) scene. The two best configurations, BVH8-RS-C and BVH4-RS-C
           achieve the lowest memory traffic consistently over multiple diffuse bounces. The other scenes exhibit similar statistics.}
\end{figure*}

Our quantization approach necessarily introduces some approximation compared to
full floating-point precision. \Cref{fig:visual-quality-bits} presents a
side-by-side comparison between images rendered using our compressed BVH and
triangle representation versus a floating-point reference. The compression
artifacts manifest primarily as discretized edges and subtle geometric
distortions, including slightly enlarged or displaced elements. These artifacts
become particularly pronounced at lower ray-precision settings, where the
quantization effects are clearly visible. As ray precision increases, we observe
a significant reduction in the blocky artifacts, effectively producing smoother
image quality that approximates the reference rendering more closely. \Cref{fig:visual-quality-pt} shows a side-by-side comparison of path traced images with and without compression. The most notable artifacts appear at edges and corners where the quantization error in the origin and direction of a bounced ray lead to different results as in the uncompressed version.

The magnitude of visual artifacts correlates strongly with the precision available at the leaf-level in the BVH. As detailed in Section 3, large leaf nodes (resulting from either high triangle counts or geometrically extensive triangles) force their scale factors to propagate upward through the hierarchy. This propagation can cause other leaf nodes to experience a decrease in precision. For example, the \textit{Viking} scene in \Cref{fig:visual-quality-bits} (bottom) requires coarse scale factors of (-3, -5, -4) for all leaf nodes using a 4-wide BVH. In contrast, the smallest leaf nodes in this scene would optimally utilize much finer scale factors of (-22, -22, -21) if evaluated independently. This substantial disparity in quantization granularity results in pronounced discretization errors for smaller triangles, explaining the visible artifacts in scenes with high geometric complexity variation. In contrast, the \textit{Teapot} scene (Fig. \ref{fig:visual-quality-bits}, top) exhibits considerably less geometric distortion, as it requires scale factors of (-4, -4, -4) at the leaf-level while the smallest leaves would only need moderately finer factors of (-10, -12, -11). This reduced quantization disparity directly translates to less perceptible visual artifacts in simpler scenes with more uniform geometric scale. It follows that preprocessing the mesh so that the size difference between the largest and smallest triangles becomes smaller directly improves the visual quality of the image computed using compression.

\section{Conclusion}
\begin{figure*}
  \centering
  \includegraphics[width=.8\linewidth]{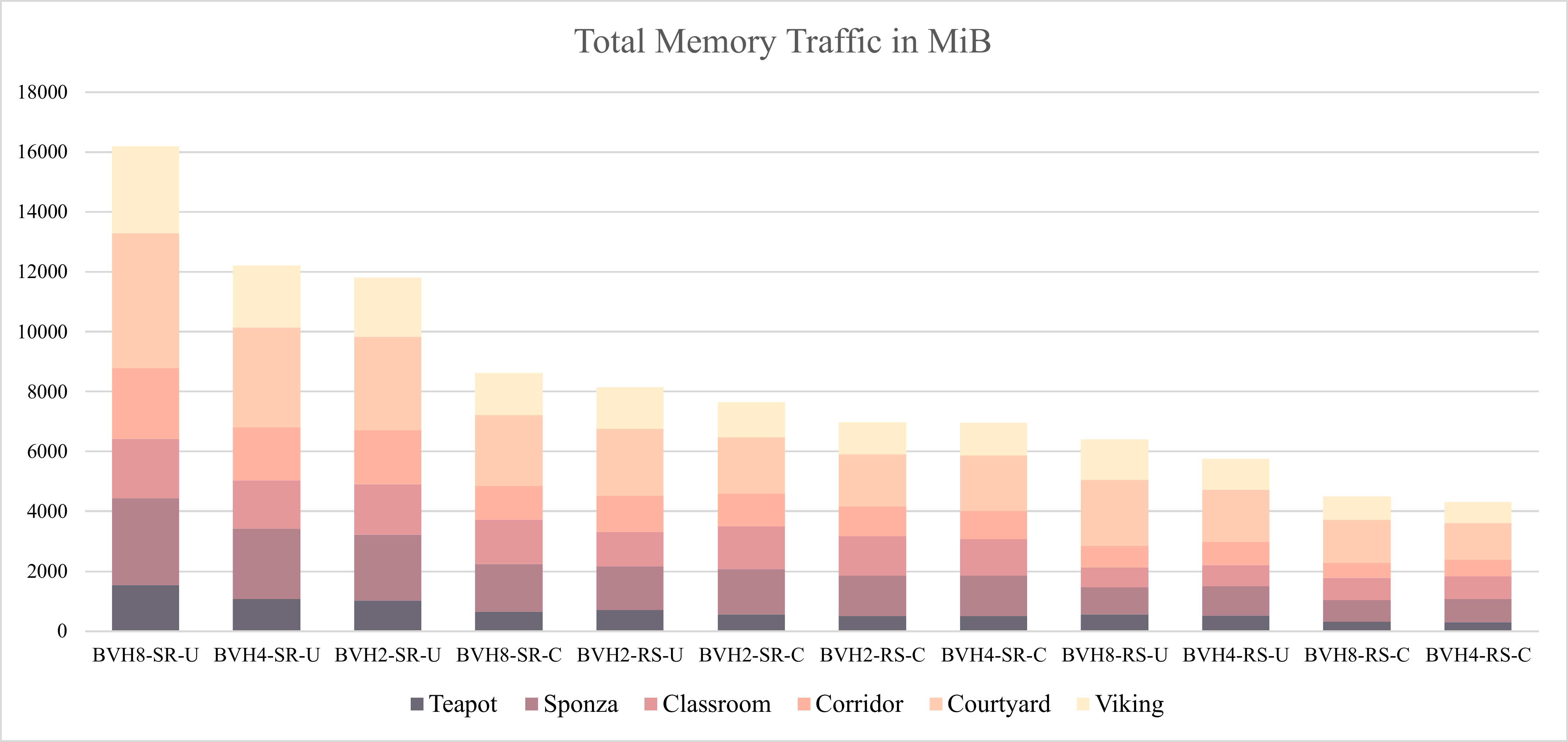}
  \caption{\label{fig:leaflevel}
           Comparison of the total memory traffic for all evaluated
           configurations. The two best configurations, BVH8-RS-C and BVH4-RS-C
           do not consistently achieve lowest memory traffic in
           all scenes (see \Cref{tab:total-traffic}), but more so for higher
           triangle counts.}
\end{figure*}

We present an approach to reduce memory loads and stores in a ray tracing core by 
systematically going through all sources of memory access: geometry bounds,
triangles, and ray traversal stacks. We leverage findings from previous work and
present an analysis of an integrated system combining. Interesting trade-offs arise in many cases: quantized bounds reduce
the absolute memory footprint of BVHs but increase the number of ray/box intersections
because the bounds grow due to rounding. Ray stream tracing can extract coherence
even in incoherent rays when accessing the bounds on the upper levels of the BVH.
Compared to single ray tracing, this introduces additional memory traffic for a combined
ray stream stack. Quantizing the triangle data can remove memory indirections and
reduce the overall memory footprint significantly. This comes, however, with a visual
impact especially for very low bit widths, and good BVH traversal should be dominated
by accessing bounds, not triangles.
From our in-depth analysis, we draw
several conclusions:
\paragraph*{Visual impact of Quantization.}
Our quantization scheme introduces slight shifts in bounding box coordinates and
triangle vertices compared to their original floating-point values. While these
shifts are imperceptible in rendered images, a direct comparison with
floating-point rendering reveals subtle geometric displacements.

\paragraph*{Scaling with scene complexity.}
Larger scenes show different trade-offs than smaller scenes, and the 
winning technique shifts from BVH8 to BVH4 as triangle and bound data
begins to dominate the ray traversal stack.

\paragraph*{BVH8 vs. BVH4.}
While BVH8 yields better results in small scenes, BVH4 reduces memory traffic the most at high
triangle counts (\Cref{tab:total-traffic}). This emphasizes the need to evaluate the algorithms on actual
workloads. Traversing a BVH8 with ray streams entails additional algorithmic complexity, with the promise to amortize
quantization data over more child boxes. From our results, we can not prove that
the added complexity in constructing 8-wide BVHs compensates the reduction in
memory traffic over BVH4. An actual hardware implementation would thus need to
carefully weigh in the improved SIMD opportunities when intersecting rays with
eight boxes at a time. Since the memory numbers of BVH4-RS and BVH8-RS are very
close, it might still be that BVH8 result in better performance due
to the utilization of wider SIMD units. Additionally, as shown by 
Ylitie et al.~\cite{ylitie_efficient_2017}, BVH8 builders can be designed with
more focus on low memory footprint than what we got from Embree.

In summary, our numbers demonstrate that we can achieve a significant
reduction in memory traffic for ray tracing
(see \Cref{fig:leaflevel}) at manageable visual impact. These findings suggest that future hardware implementations can leverage these quantization and compression techniques to optimize memory bandwidth utilization. A useful abstraction
to hide the intricacies of fixed-point arithmetic from rendering engineers
could for instance be that the internal data representation of the 
acceleration structure in the Vulkan API would be the quantized representation of
BVH and triangles. This way it would be opaque to the user and could be swapped
out by updated algorithms in future driver releases.

\subsection{Limitations and Future Work}
In this work, we focused only on the memory traffic implications of a ray
tracing core in several variants. We left some considerations out of the
analysis, which can have a big impact on performance.
\paragraph*{Mesh preprocessing.}
To preserve watertightness of the input mesh, we adjust the global
quantization gap to the largest leaf node in the scene. This means that large
triangles on the input deteriorate ray tracing accuracy. Modern geometry moves
towards smaller triangles, suitable for on-the-fly tessellation
\cite{cluster_accel2025,tamy2008,lenz,nanite2021},
so this problem might be less apparent in the future. Still, moderate subdivision
before BVH construction will improve the accuracy of our approach.
\paragraph*{BVH build times.}
We did not examine BVH construction. Preprocessing and quantizing geometry and bounds (\Cref{sec:results})
adds some overhead to the build times that might be alleviated with a more specialized
BVH build routine.
\paragraph*{Actual hardware.}
We showed a CPU simulation of low-memory ray tracing, working on
compressed primitives. To achieve performance competitive with current hardware ray tracing units, specialized hardware components must be designed to efficiently execute the required fixed-point operations.
\paragraph*{Bit widths of the fixed-point units.}
Our current implementation moves rays and triangles into a shared, high
resolution world space (64-bit precision) before intersection. This
necessitates hardware units running at high bit widths. To avoid this, we could
transform the ray origins into low precision (8-bit) leaf node space before
intersection. Since our leaf nodes all operate in the same global precision, we
do not foresee any issues transforming the ray origin to a point on the leaf node
box first. This has the potential to save a lot of die space for the arithmetic
intersection units.
\bibliographystyle{eg-alpha-doi} 
\bibliography{rtbib}       


\appendix
\section{Fixed-Point Arithmetic}
Implementing fixed-point arithmetic requires careful attention to precision and overflow. We use the following approach for key operations, taking fixed-point numbers of varying formats FixedP(integer value, R, Q).

\begin{algorithm}
\caption{FixedPoint Addition/Subtraction}\label{code:fixed-point-add}
\DontPrintSemicolon 
\KwIn{this, other: FixedP}
\KwOut{result: FixedP}
$R' \gets \max(this.R, other.R) + 1$\;

\uIf{$this.Q = other.Q$}{
    \Return{FixedP(this.val $\pm$ other.val, R', this.Q)}
}
\uElseIf{$this.Q < other.Q$}{
    $rescaled \gets this.rescale(other.Q)$\;
    \Return{FixedP(rescaled.val $\pm$  other.val, R', other.Q)}
}
\Else{
    $rescaled \gets other.rescale(this.Q)$\;
    \Return{FixedP(this.val $\pm$  rescaled.val, R', this.Q)}
}
\end{algorithm}



\begin{algorithm}
\caption{FixedPoint Multiplication}\label{code:fixed-point-multiply}
\DontPrintSemicolon 
\KwIn{this, other: FixedP}
\KwOut{result: FixedP}
\Return{\text{FixedPRQ}(this.val $\cdot$ other.val, this.R + other.R, this.Q + other.Q)}
\end{algorithm}

\begin{algorithm}
\caption{FixedPoint Division}\label{code:fixed-point-divide}
\DontPrintSemicolon 
\KwIn{this, other: FixedP}
\KwOut{result: FixedP}
$numerator \gets this.val \ll (other.Q + other.R)$\;
\Return{\text{FixedP}(numerator / other.val, this.R + other.Q, other.R + this.Q)}
\end{algorithm}


\end{document}